\documentclass{aa}
\usepackage{graphicx}
\usepackage{multirow}
\usepackage{txfonts}
\usepackage{natbib}
\usepackage{lscape}
\bibpunct{(}{)}{;}{a}{}{,}  % to follow the A&A style
\begin{document}
\title{Mapping photodissociation and shocks in the vicinity of \mbox{Sgr A$^*$}
%\subtitle{}
 \thanks{Figures 3--7 and 13 are only available in the online version via http://www.edpsciencies.org}.
 \fnmsep\thanks{Based on observations with the IRAM 30-m telescope. IRAM is supported by CNRS/INSU
         (France), the MPG (Germany) and the IGN (Spain).}
       }
\author{M. A. Amo-Baladr\'on
 \inst{1}
 \and
 J. Mart\'in-Pintado
 \inst{1}
 \and
 S. Mart\'in
 \inst{2,3}
%\fnmsep\thanks{Just to show the usage of the elements in the author field} %footnote
 }
\institute{Centro de Astrobiolog\'ia (CSIC/INTA), 
           Ctra. de Torrej\'on a Ajalvir km 4, 
           E-28850, Torrej\'on de Ardoz, Madrid, Spain\\
           \email{arancha@damir.iem.csic.es}
           \and
           European Southern Observatory, 
           Alonso de C\'ordova 3107, Vitacura, 
           Casilla 19001, Santiago 19, Chile
           \and
           Harvard-Smithsonian Center for Astrophysics,             
           60 Garden Street, Cambridge, MA 02138, USA\\
          %\thanks{The university of heaven temporarily does not accept e-mails} %footnotes
}
\date{Received ---; accepted ---}
%
% \abstract{}{}{}{}{} 
% 5 {} token are mandatory
% 
\abstract
%context heading (optional)
%{} leave it empty if necessary  
{}
%aims heading (mandatory)
{We study the chemistry in the harsh environments of galactic nuclei using the nearest one, the Galactic center (GC).} 
%
%methods heading (mandatory)
{We have obtained maps of the molecular emission within the central five arcminutes (\mbox{12 pc}) 
of the GC in selected molecular tracers: \mbox{SiO(2--1)}, \mbox{HNCO(5$_{0,5}$--4$_{0,4}$)}, and the
\mbox{$J =$ 1 $\rightarrow$ 0} transition of \mbox{H$^{13}$CO$^+$}, \mbox{HN$^{13}$C}, and \mbox{C$^{18}$O}
at an angular resolution of 30$''$ (\mbox{1.2 pc}).
The mapped region includes the circumnuclear disk (CND) and the two surrounding giant 
molecular clouds (GMCs) of the \mbox{Sgr A} complex, known as the 20 and 
\mbox{50 km s$^{-1}$} molecular clouds.
Additionally, we simultaneously observed the \mbox{$J =$ 2 $\rightarrow$ 1} and \mbox{$J =$ 3 $\rightarrow$ 2} 
transitions of \mbox{SiO} toward selected positions to estimate the physical conditions of the molecular gas 
using the large velocity gradient approximation.}
%
%results heading (mandatory)
{The \mbox{SiO(2--1)} emission shows all the molecular features identified in previous studies, 
covering the same velocity range as the \mbox{H$^{13}$CO$^+$(1--0)} emission, which also 
presents a similar distribution.  
In contrast, \mbox{HNCO(5--4)} emission appears in a narrow velocity range mostly concentrated
in the 20 and \mbox{50 km s$^{-1}$} GMCs. A similar trend follows the \mbox{HN$^{13}$C(1--0)} emission.
The HNCO column densities and fractional abundances present the highest contrast, with difference factors
of \mbox{$\geq$60} and 28, respectively. Their highest values are found toward the cores of
the GMCs, whereas the lowest ones are measured at the CND.
SiO abundances do not follow this trend, with high values found toward the CND, as well as the GMCs.
By comparing our abundances with those of prototypical Galactic sources we conclude that HNCO, similar
to SiO, is ejected from grain mantles into gas-phase by nondissociative C-shocks. This results
in the high abundances measured toward the CND and the GMCs.
However, the strong UV radiation from the Central cluster utterly photodissociates HNCO 
as we get closer to the center, whereas SiO seems to be more resistant against UV-photons 
or it is produced more efficiently by the strong shocks in the CND.
This UV field could be also responsible for the trend found in the \mbox{HN$^{13}$C} abundance.}
%
%conclusions heading (optional), leave it empty if necessary 
{We discuss the possible connections between the molecular gas at the CND and the GMCs 
using the \mbox{HNCO/SiO}, \mbox{SiO/CS}, and \mbox{HNCO/CS} intensity ratios as probes 
of distance to the Central cluster.
In particular, the \mbox{HNCO/SiO} intensity ratio is proved to be an excellent tool for evaluating the 
distance to the center of the different gas components.}
\titlerunning{Mapping the surrounding of \mbox{Sgr A$^*$}}
\authorrunning{M. A. Amo-Baladr\'on et al.}
\maketitle
%
%_____________________________________________________________________________________________________________________________
%
\section{Introduction}
\label{intro}
\begin{figure}          
\centering    
\includegraphics[angle=0,width=8cm]{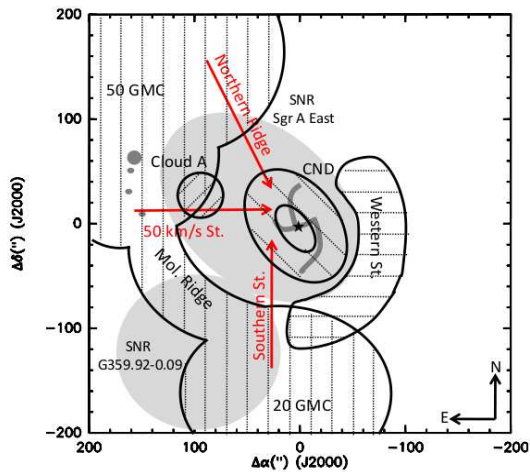}
\caption{Sketch of the 12 central parsecs of the Galaxy. Gray regions represents
radio continuum features. Big ellipses correspond to the SNRs \mbox{Sgr A} East and \mbox{G 359.92$-$0.09},
the central minispiral to the \ion{H}{ii} region \mbox{Sgr A} West, and the four little circles to
the \mbox{Sgr A} East compact \ion{H}{ii} regions. The central star represents \mbox{Sgr A$^*$}. 
Molecular gas features are represented by stripped regions, whereas red arrows toward the center point out 
the possible connections between the CND (the elliptic disk) and the CMCs of this region. 
See the text for more details.
\label{fig1}}  
\end{figure}

Its proximity (\mbox{8.0$\pm$0.5 kpc}; \citealt{Rei93}) makes the Galactic center (GC) the best 
laboratory for understanding the heating and chemistry of the interstellar medium (ISM) in the harsh environment 
of galaxy nuclei.
Just in its central \mbox{5 arcmin} (12 pc), one can find a massive black hole surrounded by a rotating 
circumnuclear disk (CND) of dust and gas, \ion{H}{ii} regions, massive stellar clusters, two supernova remants (SNRs), 
and two giant molecular clouds (GMCs). All these features coexist, interact, and can be spatially resolved at a resolution
currently impossible even in the nearest extragalactic nucleus.\\
\indent At its dynamical center, the GC harbors a massive black hole of \mbox{$\sim$4.0 $\times$ 10$^6$ M$_\odot$} 
\citep{Ghe05} that is associated with the strong compact source of nonthermal emission \mbox{Sgr A$^*$},
visible in radio continuum maps \citep{YZM87}.
Converging to \mbox{Sgr A$^*$}, there are arc-shaped ionized gas streamers with a minispiral morphology (\mbox{Sgr A West})
that seem to be feeding the nucleus \citep{LoC83,RoG93}. Both the black hole and \mbox{Sgr A West} are surrounded by a 
ring of molecular gas and dust (the CND; \citealt{Bec82,Gus87}), whose inner edge suffers UV-photoionization \citep{RoG93} 
due to a dense stellar population of massive stars in the Central cluster \citep{Kra95,Fig08}.\\
\indent The CND has been observed from radio to infrared wavelengths 
in molecular gas tracers, continuum, and atomic species (see \mbox{Table \ref{tab1}} for references).
These studies have characterized the CND as dense 
(\mbox{$\sim$10$^{5-6}$ cm$^{-3}$}; \citealt{Mar93,Mas95}), clumpy, and very turbulent 
(with large linewidths). It extends from its inner edge, \mbox{1.6 pc} from \mbox{Sgr A$^*$}, 
to \mbox{$\ga$2 pc}, according to interferometric studies \citep{Wri01,Chr05}, and it has
an inclination (rotation axis from the line of sight; LOS) of about \mbox{60$\degr$--70$\degr$}, 
with the major axis aligned approximately along the Galactic plane 
(position angle of \mbox{$\sim$30$\degr$} east of north; \citealt{Gus87,Jac93,Mas95}).
The gas is moving in circular orbits, rotating around 
the nucleus with a velocity of \mbox{110 km s$^{-1}$} \citep{LBH85,Mar93}, but it also presents noncircular 
motions (e.g.: \citealt{Gus87,Jac93,Mas95,Shu04,Chr05}). 
The CND does not appear to be an equilibrium structure. It seems to consist of several separated streamers 
in rotation around the nucleus \citep{Jac93,Wri01}. 
Virial analysis suggests that the CND could be composed of high-density clumps 
(\mbox{$\sim$10$^{6-7}$ cm$^{-3}$}) that could withstand the tidal shear making the CND   
a long-lived structure \citep{Jac93,Shu04,Chr05,MCa09}. However, \citet{Gus87} derived significantly lower densities
(\mbox{$\sim$10$^5$ cm$^{-3}$}) from their interferometric mapping of HCN, which are consistent with the densities
estimated from warm atomic gas \citep{Gen85} and CO multitransitional observations \citep{Har85}.\\  
%
%_____________________________________________________________________________________________________________________________
%
\begin{table}
\caption{Observations of the CND}
\label{tab1}
\centering
\begin{tabular}{ll} 
\hline\hline 
Tracers & References \\
\hline
H$_2$                          & \citet{Gat86} \\
                               & \citet{YZa01} \\
                               &\\
CO(1--0)                       & \citet{LBH85} \\
                               & \citet{Ser86} \\
                               & \citet{Lis95} \\
$^{13}$CO and C$^{18}$O(2--1)  & \citet{Zyl90} \\
CO(3--2)                       & \citet{Sut90} \\
                               & \citet{Den93} \\
                               & \citet{Lis95} \\
CO(7--6)                       & \citet{Bra05} \\
                               &\\
CS(2--1)                       & \citet{Ser86} \\
C$^{34}$S(2--1)                & \citet{Zyl90} \\
CS(7--6)                       & \citet{MCa09} \\
                               &\\
HCN(1--0)                      & \citet{Gus87} \\
                               & \citet{Mar93} \\
                               & \citet{Wri01} \\
                               & \citet{Chr05} \\
HCN(3--2)                      & \citet{Jac93} \\
                               & \citet{Mas95} \\
HCN(4--3)                      & \citet{Mas95} \\
                               & \citet{MCa09} \\
H$^{13}$CN(1--0)               & \citet{Mar93} \\
                               &\\
HCO$^+$(1--0)                  & \citet{Mar93} \\
                               & \citet{Wri01} \\
                               & \citet{Chr05} \\
                               & \citet{Shu04} \\
                               &\\
NH$_3$(1,1) (2,2)              & \citet{CoH99,CoH00} \\
NH$_3$(1,1) (2,2) (3,3)        & \citet{McG01} \\
NH$_3$(6,6)                    & \citet{HeH02} \\
                               &\\
Continuum at                   & \citet{Mez89} \\
mm and sub-mm                  & \citet{Zyl90} \\
wavelengths                    & \citet{Dav92} \\
                               & \citet{Den93} \\
                               & \citet{Tel96} \\
                               &\\
Atomic species                                         & \citet{LBH85}\\
(\ion{H}{i},[\ion{O}{i}],[\ion{C}{i}],[\ion{C}{ii}])   & \citet{Gen85} \\ 
                                                       & \citet{Lug86} \\
                                                       & \citet{Jac93} \\
                                                       & \citet{Ser94} \\
\hline
\end{tabular}
\end{table}
%
%_____________________________________________________________________________________________________________________________
%
\indent One question that remains without a conclusive answer is how the CND is fed. 
Since this structure is not just a transient feature losing material through infall 
(probably via \mbox{Sgr A} West, \citealt{MCa09}), there must be gas feeding the disk.
Two GMCs that lie near the GC are the best candidates to feed the CND: 
the \mbox{20 km s$^{-1}$} and the \mbox{50 km s$^{-1}$} GMCs (\mbox{M$-$0.13$-$0.08} and \mbox{M$-$0.02$-$0.07},
respectively; \citealt{GWP81,GuH83}). 
Both are connected by a ridge of compressed dust and gas that warps around the SNR
\mbox{Sgr A East}. This SNR lies behind or contains \mbox{Sgr A West} and appears as an expanding 
shell of synchrotron emission in radio continuum maps \citep{YZM87,Ped89}. 
But it is not the only SNR in the region. \mbox{G 359.92$-$0.09}, located to the south of \mbox{Sgr A East} 
(centered at \mbox{$\Delta\alpha$ =120$''$}, \mbox{$\Delta\delta$ =$-$180$''$} with respect to \mbox{Sgr A$^*$}), 
is expanding into \mbox{Sgr A East} producing its concave southeastern edge \citep{CoH00}.
Both SNRs seem to interact with the surrounding material pushing it away from or toward the 
nucleus. As a result, several streamers of gas and dust that could be feeding the CND have 
been formed \citep{Lee08}.\\  
\indent Several studies at different wavelengths have attempted to establish the relation between all 
the components in the \mbox{Sgr A} complex and their 3-dimensional (3D) location (see \citealt{CoH00,HeH05,Lee08}, 
and references therein). 
For the sake of clarity, we show in \mbox{Fig. \ref{fig1}} a 2D sketch of the region with all the previous 
identified components, following the notation of \citet{HeH05} for the molecular features in most of the cases.
These studies have resulted in the following picture:
\begin{enumerate}
\item The \ion{H}{ii} region \mbox{Sgr A West}, also known as the Minispiral, is located in front of or inside 
the SNR \mbox{Sgr A East} \citep{YZM87,Ped89}. 
Moreover, \citet{Mae02} place \mbox{Sgr A$^*$}, \mbox{Sgr A West}, and the CND (hereafter the nuclear region) 
just inside the leading edge of \mbox{Sgr A East}.
\item The SNR \mbox{Sgr A East} is pushing the \mbox{50 km s$^{-1}$} GMC both to the east and behind the SNR
along the LOS \citep{CoH00,Par04}.
\item The \mbox{20 km s$^{-1}$} GMC is located in front of the nucleus along the LOS \citep{GuD80,Par04} and
\mbox{Sgr A East} could also be expanding into this GMC \citep{Mez89}.
\item Both GMCs are connected by the Molecular Ridge, whose northern part
is farther away from the Sun along the LOS than the southern part \citep{CoH00}. 
\citet{Lee08} claim that the SNR \mbox{Sgr A East} could interact with the northern part of the ridge 
nearly perpendicular to the LOS, locating the northern part of the ridge slightly behind this SNR.
\item The SNR \mbox{G 359.92$-$0.09} is interacting with the southern part of the Molecular Ridge, the eastern edge of
the \mbox{20 km s$^{-1}$} GMC, and the southern edge of \mbox{Sgr A East} \citep{CoH00}.
\item The Western Streamer, a ridge of emission seen in NH$_3$ bordering the western edge of \mbox{Sgr A East},
is highly inclined with respect to the LOS, and it is expanding outward as does this SNR \citep{McG01,Lee08}.
\item The CND could be connected with the GMCs through three molecular gas streamers: the Southern Streamer 
\citep{Oku91}, the Northern Ridge \citep{McG01}, and the \mbox{50 km s$^{-1}$} Streamer \citep{Szc91}.
\item Finally, high-velocity \mbox{C$^{18}$O(2$-$1)} emission between \mbox{$-$80} and 
\mbox{20 km s$^{-1}$} located in projection close to the nuclear region have been identified by \citet{Gen90}.
In our maps, this emission appears as a small blue-shifted molecular cloud that we will refer it to as Cloud A.
\end{enumerate}
%
%_____________________________________________________________________________________________________________________________
%
%
%_____________________________________________________________________________________________________________________________
%
\indent The main heating mechanism in this complex region (CND and its central cavity) seems to be a combination 
of photoelectric heating by UV photons and mechanical heating by dissipation of supersonic turbulence in shocks 
\citep{Bra05,MCa09}. 
Theoretical models for the formation of the CND and the Northern Arm of the ionized Minispiral invoke infalling clouds as 
a result of cloud-cloud collisions and predict frequent collisions between clumps that would produce shock heating 
\citep{San98,VoD02}. 
Direct probes of shocks in the CND are the detections of \mbox{1720 MHz} OH masers 
with extreme velocities (\mbox{$\pm$130 km s$^{-1}$}) at its lobes reported by \citet{SjP08}.
These authors claim that this type of maser is collisionally pumped and originates in postshock regions.
\citet{Bra05}, modeling CO rotational lines, reached the conclusion that UV heating alone is not able 
to match the CO luminosities, and therefore mechanical energy dissipated into the gas in low-velocity (\mbox{10--20 km s$^{-1}$})
C-shocks with magnetic fields of \mbox{0.3--0.5 mG} could be the primary source of energy for the bulk of the 
heating of the CND material. Additional probes of shocks and/or UV radiation come from prominent H$_2$ emission
found in the CND, in its southwest and northeast lobes, just where \mbox{HCN(1--0)} emission is stronger \citep{Chr05}.
Moreover, there could be another source of shocks in the CND. In their study of the 
possible interactions between \mbox{Sgr A East} and the molecular gas features of the nuclear region, \citet{Lee08} have 
found that the southern part of the CND and its northwestern region could lie in front of the SNR \mbox{Sgr A East} 
and could be pushed toward us.
In fact, several \mbox{1720 MHz} OH masers have been detected tracing the interaction of this SNR with
the surrounding GMCs \citep{SjP08}.\\
\indent Therefore, the energetics of the central region seem to be dominated by UV radiation and 
magnetohydrodynamical shocks. 
Chemical complexity can potentially be used to locate the different molecular components along the LOS
in the central region by measuring the effects of photodissociation produced by the UV radiation from 
the Central cluster.
It has been proposed that \mbox{SiO}, \mbox{HNCO}, and \mbox{CS} are among the best molecular tracers of the 
different ISM heating mechanisms.
All of them have high critical densities suitable for GC physical conditions, so they trace similar gas 
components, but each species is sensitive to different heating mechanisms through their particular chemistry. 
\mbox{SiO} is a well-known shock tracer \citep{MPi92}, \mbox{HNCO} seems to be also associated to grain chemistry 
but it is strongly affected by photodissociation \citep{Mar08,Mar09}, whereas \mbox{CS} is marginally enhanced 
in UV \citep{Goi06,Mar08} and shocked environments \citep{Req06}.
The \mbox{SiO/CS} and \mbox{HNCO/CS} ratios can then be used to trace shocks in well-shielded regions by the enhancement 
of the SiO and HNCO abundances with respect to that of CS. In unshielded regions illuminated by UV radiation and with 
strong shocks, one expects to measure an enhancement of the \mbox{SiO/CS} ratio and the decrease in the 
\mbox{HNCO/CS} ratio.\\

\indent In this paper we present spectral imaging of the molecular emission from the central region of the GC
(12 pc), which includes the CND and two GMCs of the \mbox{Sgr A} complex.
The selected molecular lines were the \mbox{$J =$ 2 $\rightarrow$ 1} transition of SiO,
the \mbox{$J =$ 5$_{0,5} \rightarrow$ 4$_{0,4}$} transition of \mbox{HNCO} (hereafter \mbox{HNCO(5--4)}),
and the \mbox{$J =$ 1 $\rightarrow$ 0} transition of \mbox{H$^{13}$CO$^+$}, \mbox{HN$^{13}$C}, and \mbox{$^{18}$CO}. 
In \mbox{Sect. \ref{Obs}} the observation and data reduction procedures are described.
\mbox{Sect. \ref{Res}} presents the data for each observed species in selected velocity ranges for better
comparison with radio continuum maps. Kinematics is described with declination-velocity maps at constant 
right ascension intervals. 
Physical conditions at selected positions, as well as column densities and fractional abundances, 
are derived in \mbox{Sects. \ref{PhysConds}} and  \ref{FracAbun}.
Finally, in \mbox{Sect. \ref{Dis}} the changes found in molecular abundances are discussed,
compared to prototypical sources, and used to probe the different heating mechanisms in 
the region and to establish the location of the different molecular components in the nuclear region 
relative to the Central cluster.\\
%
%_____________________________________________________________________________________________________________________________
%
\section{Observations} 
\label{Obs}
The observations were carried out with the \mbox{IRAM-30m} radiotelescope at 
Pico Veleta (Spain) during the summer of 2006.
The mapping of the lines were made simultaneously by tuning the two \mbox{3 mm} receivers (A and B) 
to two different frequencies: \mbox{86.892} and \mbox{109.905 GHz}. 
Within the first frequency band we observed the \mbox{SiO(2--1)}, 
\mbox{H$^{13}$CO$^+$(1--0)}, and \mbox{HN$^{13}$C(1--0)}
lines, whereas within the second frequency band
we observed the \mbox{HNCO(5--4)} and \mbox{C$^{18}$O(1--0)} lines.
Both receivers were tuned to a single sideband with image rejections of
\mbox{30 dB} and \mbox{25 dB}. 
The system temperatures ranged between \mbox{154--176 K} and \mbox{222--265 K} during 
the observations for the lower and higher frequency setups, respectively. 
We used filterbanks (\mbox{512 $\times$ 1 MHz}) as spectrometers, providing a spectral 
resolution of \mbox{3.4 km s$^{-1}$} at \mbox{86.9 GHz}, and \mbox{2.7 km s$^{-1}$} at 
\mbox{109.9 GHz}, enough to resolve the wide emission lines of the GC (\mbox{$\sim$20 km s$^{-1}$}).\\
\indent Data were taken using the on-the-fly (OTF) technique
with a telescope dump time (integration time on each position) of \mbox{1 s} 
and was fully sampled at \mbox{86.9 GHz}. 
The maps cover an area of \mbox{300$''$ $\times$ 300$''$} 
(\mbox{12 $\times$ 12 pc} at the GC distance of \mbox{8.0 kpc}) centered on 
the position of the radio source \mbox{Sgr A$^*$}
(\mbox{($\alpha$, $\delta$)$_{\rm {J2000}}$= 17$^{\rm d}$45$^{\rm m}$40$\fs$031, $-$29$\degr$00$'$28$\farcs$58}).
We made eight coverages of the region alternating the scanning direction 
between right ascension and declination. The orthogonal covers were made in such a way that
the \mbox{$\sim$200$''$ $\times$ 200$''$} central region was finally mapped in both 
directions, in order to minimize possible scanning artifacts.
The half-power beamwidth (\mbox{HPBW}) of the telescope was 28$''$ at the 
frequency of \mbox{86.9 GHz}, and 22$''$ at \mbox{109.9 GHz}.\\
\indent Spectra were calibrated using the standard dual load system. 
We used the antenna temperature scale (\mbox{$T_{\rm A}^*$}) for the line intensities because
the emission is rather extended and fills the beam.
We used \mbox{($l, b$)=($-$0.02$\degr$, $-$0.20$\degr$)} as reference position,
close enough to minimize the time lost due to telescope switching and differences 
in atmospheric airmass. This position shows almost no emission in the \mbox{CS(1--0)} 
survey of \citet{Tsu99}. 
To confirm that the reference was free of emission, it was checked against another 
further reference (\mbox{($l, b$)=($-$0.25$\degr$, $-$0.25$\degr$)}). 
The reference position shows no emission at a noise level of 12 and \mbox{23 mK} for the 
low and high frequency spectra, respectively.
A focus check was performed at the beginning of each observing session, and pointing was checked
at the beginning and in the middle of each observing session.\\
\indent Data reduction was done with the \texttt{GILDAS} package.
The reduction process was as follows.
First, bad channels and standing waves were removed from the spectra, and
typical baselines up to order 3, in the case of spectra at \mbox{86.9 GHz}, 
and up to order 5, at \mbox{109.9 GHz}, were subtracted. 
In the case of the high-frequency spectra, a baseline of order 0 would had been enough, 
but a baseline of higher degree was needed to correct the baseline at the backend edges. 
We also used the CS data cube of the \mbox{$J =$ 1 $\rightarrow$ 0} line taken by \citet{Tsu99}. 
To make all our molecular line cubes and that of \mbox{CS(1--0)} comparable, our spectra were 
resampled to the same velocity resolution as for the \mbox{CS(1--0)} cube (\mbox{5 km s$^{-1}$}).
Finally, we constructed all the molecular line maps with an \mbox{HPBW} of 30$''$ 
(the beam of the \mbox{86.9 GHz} and \mbox{CS(1--0)} cubes).  
The final data cubes have an rms noise per channel of 27, 37, and \mbox{100 mK}
for the \mbox{SiO}/\mbox{H$^{13}$CO$^+$}/\mbox{HN$^{13}$C}, \mbox{HNCO}/\mbox{C$^{18}$O}, 
and \mbox{CS} lines, respectively.\\
\indent We also present observations of the \mbox{$J =$ 2 $\rightarrow$ 1} and \mbox{$J =$ 3 $\rightarrow$ 2} 
transitions of \mbox{SiO} taken simultaneously at selected positions. 
The \mbox{HPBW} of the telescope at the rest frequency of the \mbox{SiO(3--2)} transition (\mbox{130.26861 GHz}) 
was 19$''$. The system temperature was around 150 for the \mbox{(2--1)} and \mbox{270 K} for \mbox{(3--2)} 
transitions, resulting in an rms noise \mbox{$\leq$32 mK}.
The \mbox{(40$''$, $-$120$''$)} position was observed in a past campaign and presents system temperatures of 
278 and \mbox{423 K}, and rms noise values of 56 and \mbox{80 mK} for the \mbox{(2--1)} and \mbox{(3--2)} 
transitions.\\ 
%
%_____________________________________________________________________________________________________________________________
%
\section{Results}  
\label{Res}
%
%_____________________________________________________________________________________________________________________________
%
\begin{figure*}
\centering
\includegraphics[width=16cm]{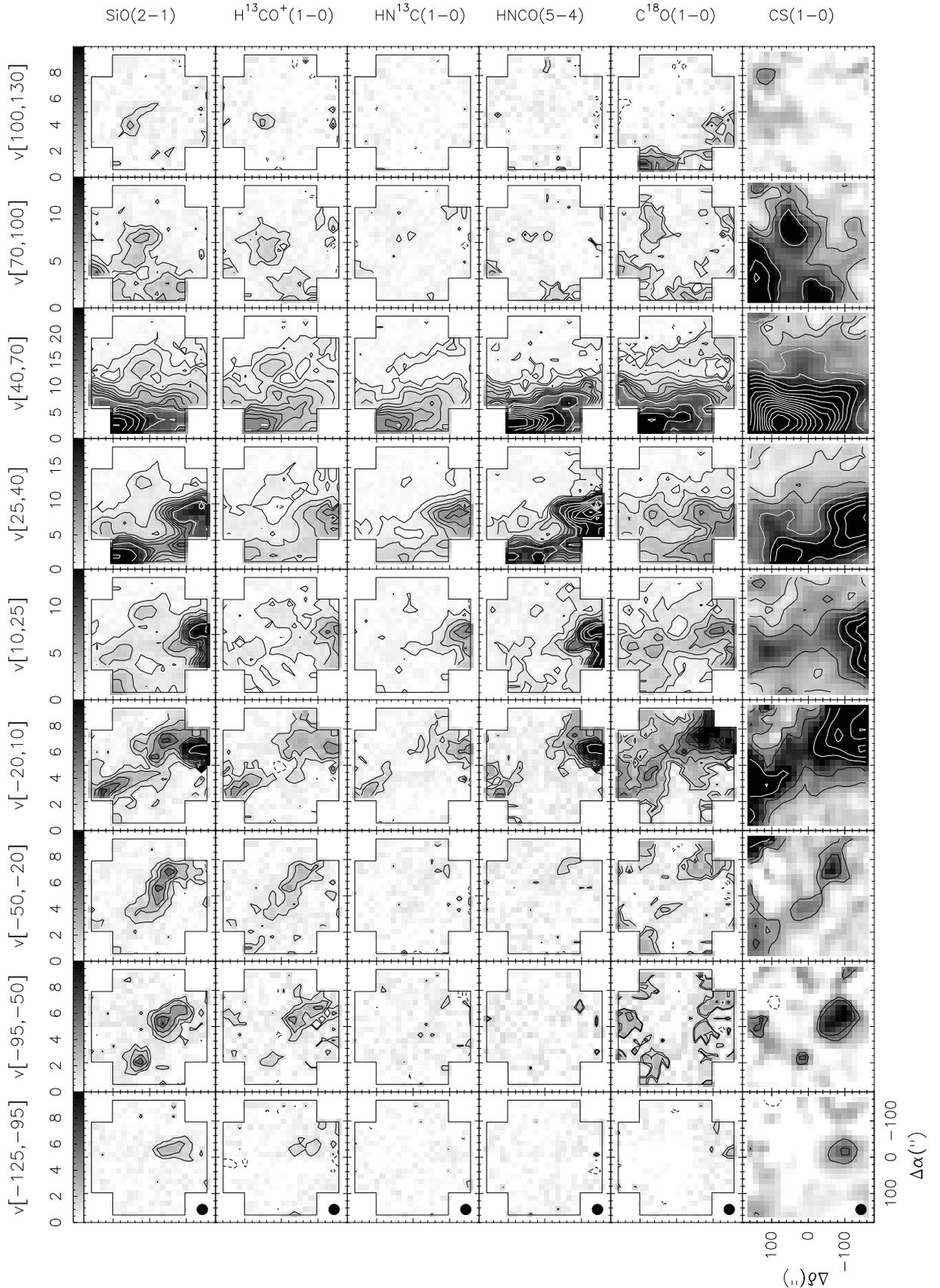}
\caption{\scriptsize{Selected velocity-integrated maps for the different molecules observed at the 
\mbox{Sgr A} complex.
Maps of the \mbox{CS(1--0)} line \citep{Tsu99} are displayed at the bottom. 
Each column corresponds to the same velocity range, indicated at the top of each column.
The wedge at the top shows the intensity gray scale for each column. 
Contour levels for the integrated-velocity emission (in the \mbox{$T_{\rm A}^*$} scale) of all the molecular lines, 
except \mbox{CS(1--0)}, are \mbox{$-$3$\sigma$} (dashed contour), \mbox{3$\sigma$}, from 2.0 to 12 in steps of  
\mbox{1.5 K km s$^{-1}$} and from 12 in steps of \mbox{6 K km s$^{-1}$}.
Contour levels for the velocity-integrated \mbox{CS(1--0)} emission are \mbox{$-$3$\sigma$} (dashed contour), 
\mbox{3$\sigma$}, and from 6 in steps of \mbox{6 K km s$^{-1}$}.
White contours are from \mbox{12 K km s$^{-1}$}.
The \mbox{3$\sigma$} levels corresponding to the 15, 30, and \mbox{45 km s$^{-1}$} wide velocity ranges are 
0.7, 1.0, and \mbox{1.2 K km s$^{-1}$} for the \mbox{SiO(2--1)}, \mbox{H$^{13}$CO$^+$(1--0)}, and \mbox{HN$^{13}$C(1--0)}
maps; 1.0, 1.3, and \mbox{1.7 K km s$^{-1}$} for the \mbox{HNCO(5--4)} and \mbox{C$^{18}$O(1--0)} maps; and 
2.6, 3.7, and \mbox{4.5 K km s$^{-1}$} for the \mbox{CS(1--0)} maps.
The beam size (30$''$) is shown in the bottom-left corner of the first column panels.
\mbox{Sgr A$^*$} is the origin for the offset coordinates.}
\label{fig2}}  
\end{figure*}
%
%_____________________________________________________________________________________________________________________________
%
\begin{figure*}
\centering
\includegraphics[width=14cm]{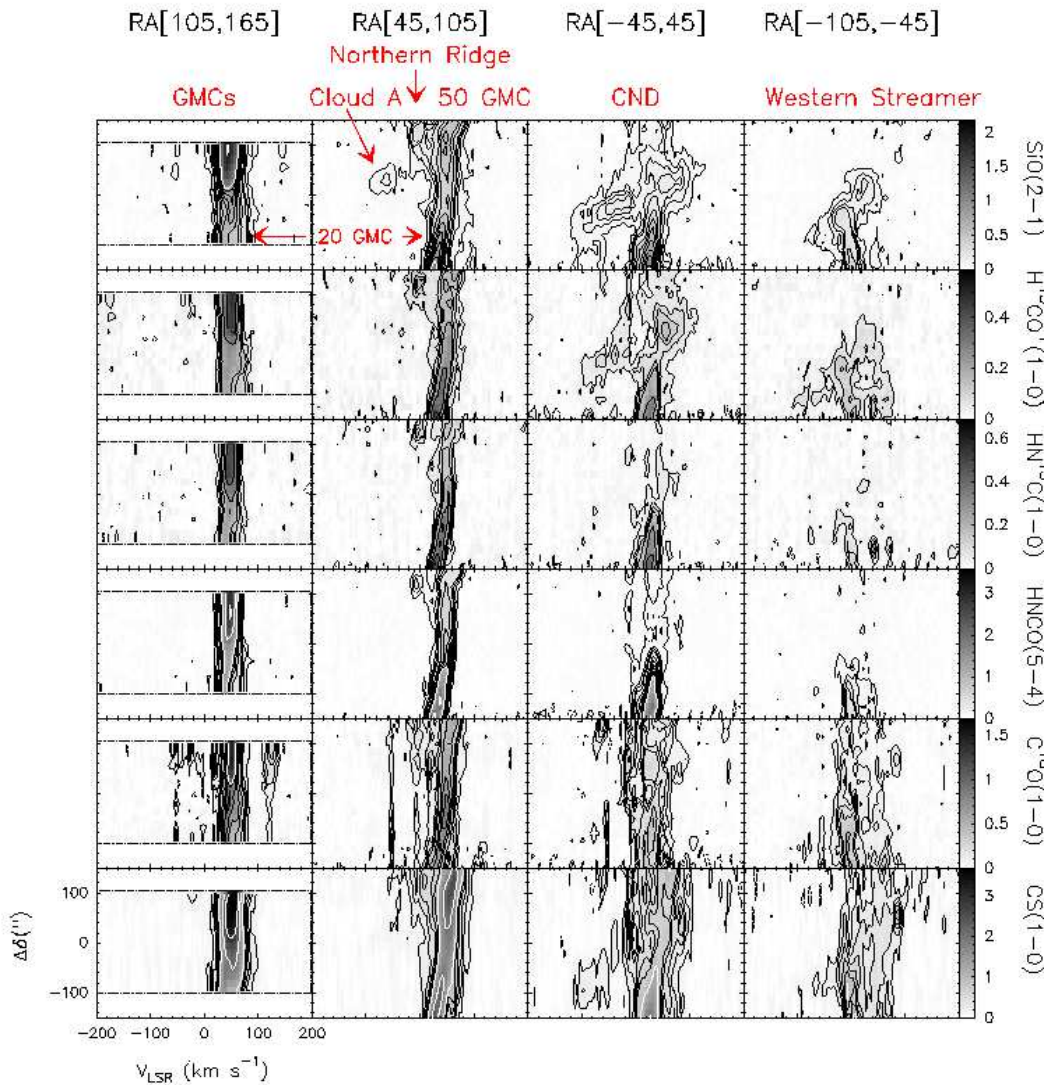}
\caption{\mbox{Declination-velocity} maps for the different molecules 
observed at the \mbox{Sgr A} complex.
The \mbox{$\Delta\delta$-v} maps have been averaged in the right ascension ranges 
shown at the top of each column.
Black contour levels are \mbox{$-$3} (dashed contour), 3, 6, 9, and \mbox{12$\sigma$}. 
The following contours are in steps of \mbox{9$\sigma$} until \mbox{1 K}. 
In the case of the \mbox{CS(1--0)} maps, black contour levels are \mbox{$-$3} (dashed contour), 3, 6, 9, 12, and 
\mbox{15$\sigma$}. The following contours are in steps of \mbox{12$\sigma$}.  
White contours begin from \mbox{1 K} and are in steps of \mbox{1 K}. 
The \mbox{3$\sigma$} levels corresponding to the \mbox{60$''$} and \mbox{90$''$} wide right ascension ranges are 
0.04 and \mbox{0.03 K} for the \mbox{SiO(2--1)}, \mbox{H$^{13}$CO$^+$(1--0)}, and \mbox{HN$^{13}$C(1--0)}
maps; 0.06 and \mbox{0.04 K} for the \mbox{HNCO(5--4)} and \mbox{C$^{18}$O(1--0)} maps; and 
0.15 and \mbox{0.12 K} for the \mbox{CS(1--0)} maps.
The wedge on the right shows the intensity gray scale for each molecular line.
\label{fig3}}  
\end{figure*}
%
%_____________________________________________________________________________________________________________________________
%
\begin{figure*}
\centering
\includegraphics[width=10cm]{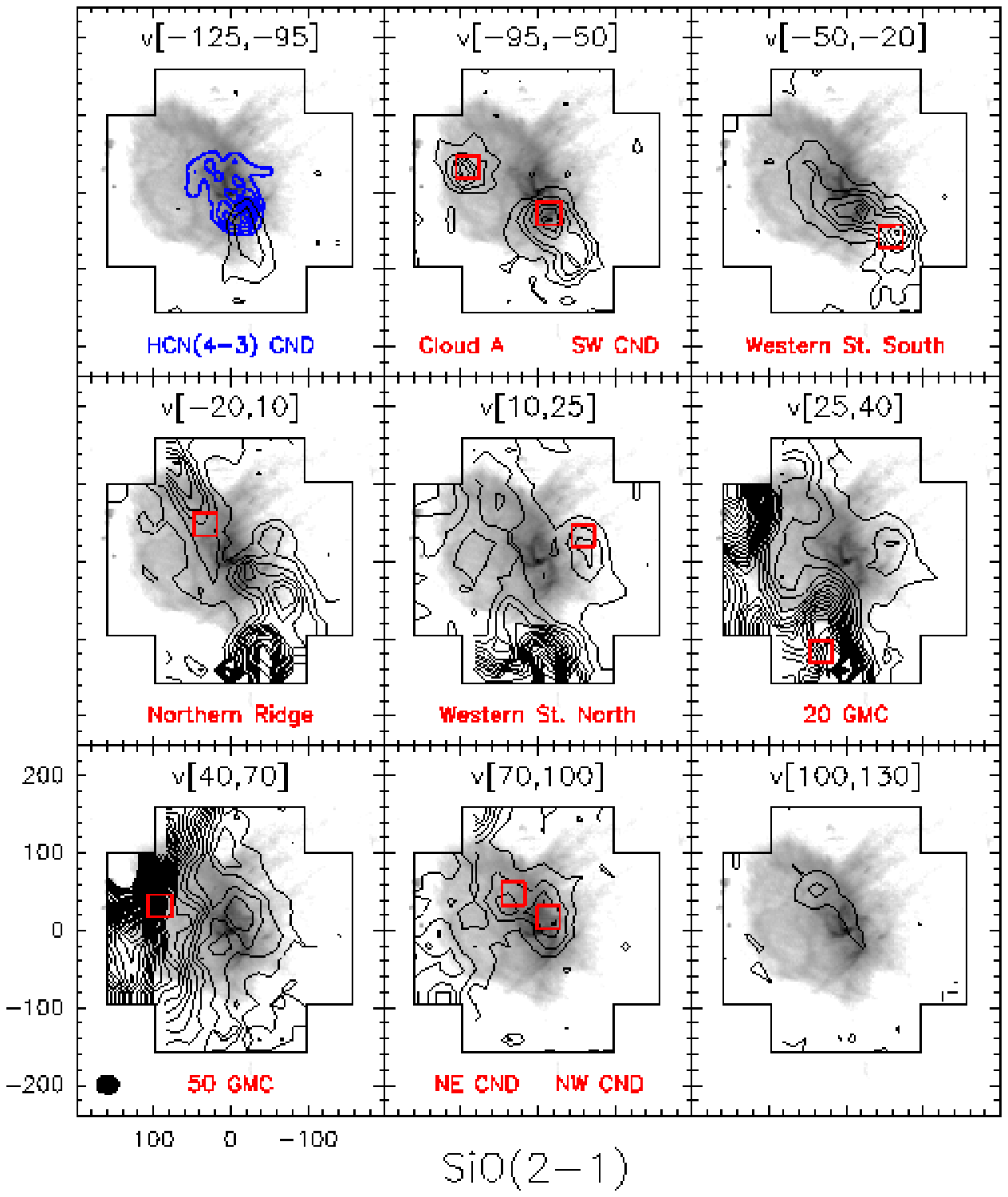}
\caption{Selected velocity-integrated maps of \mbox{SiO(2--1)} emission superimposed on 
the radio continuum image of \citet{YZM87} at \mbox{6 cm} in gray scale. 
We can appreciate the thermal feature \mbox{Sgr A} West, also known as the Minispiral, 
in the center of the image. The first two level contours are \mbox{$-$0.7} 0.7, \mbox{$-$1.0} 1.0, and
\mbox{$-$1.2} \mbox{1.2 K km s$^{-1}$} for integrated velocities 15, 30, and \mbox{45 km s$^{-1}$},
respectively. Steps between levels are \mbox{1 K km s$^{-1}$}.
Open squares indicate the spatially integrated areas where spectra 
of \mbox{Fig. \ref{fig6}} were obtained, which are associated with the main identified molecular 
features, whose names are indicated at the bottom of each corresponding panel. 
In the first panel, we also show the \mbox{HCN(4--3)} map of the CND of \citet{MCa09} smoothed to 15$''$
in thick blue contours.
\label{fig4}}  
\end{figure*}
%
%_____________________________________________________________________________________________________________________________
%
\begin{figure*}
\centering
\includegraphics[width=12cm]{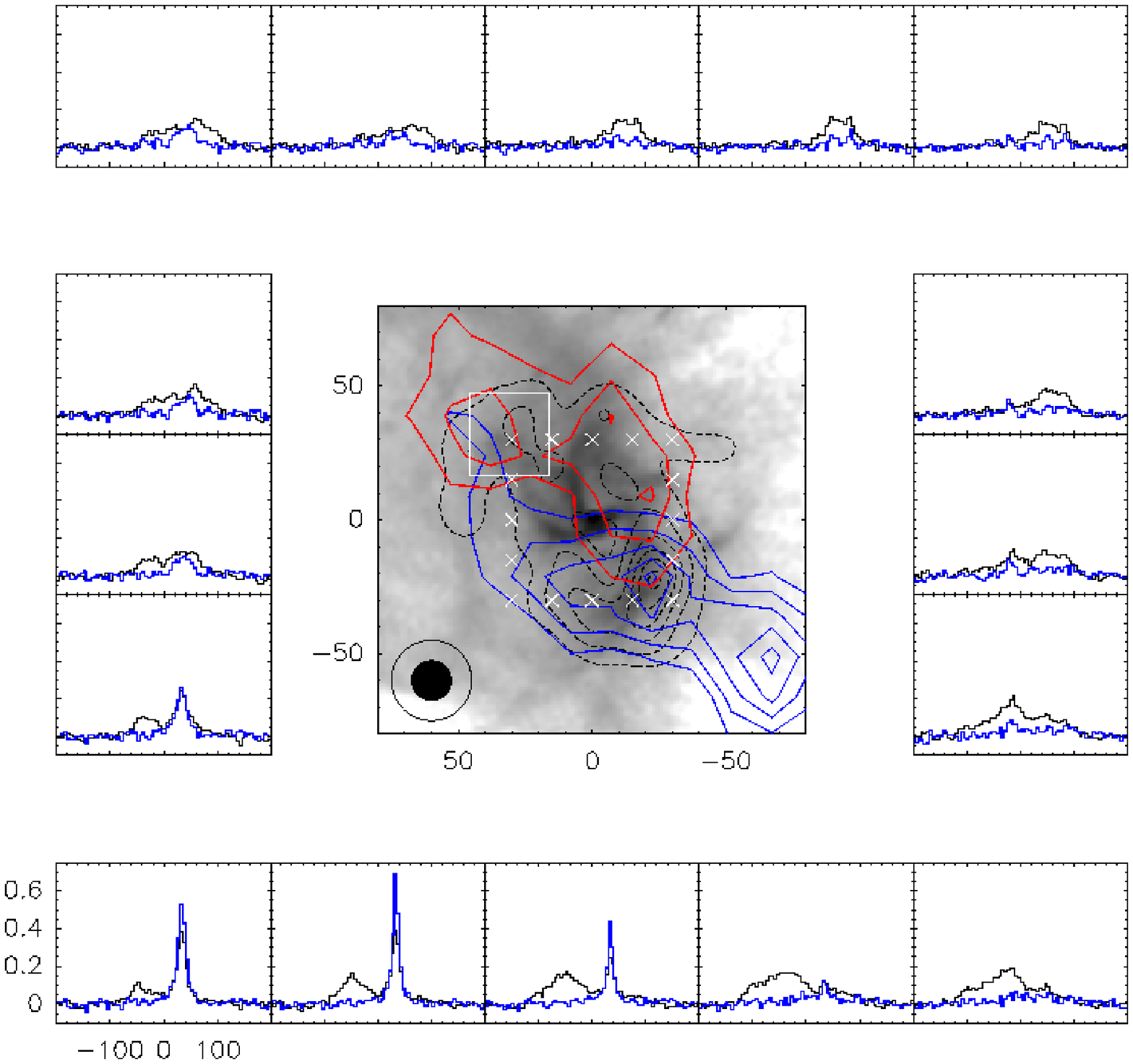}
\caption{Composition of the velocity-integrated maps of the \mbox{SiO(2--1)} emission at extreme velocities
(velocity ranges between \mbox{[$-$50, $-$20] km s$^{-1}$} in thin blue contours and \mbox{[70, 100] km s$^{-1}$} in thick red ones),
superimposed on the radio continuum image of \citet{YZM87} at \mbox{6 cm} in gray scale.
Black dashed contours correspond to the \mbox{HCN(4--3)} interferometric map of the CND \citep{MCa09} smoothed to \mbox{15$''$} resolution.
The \mbox{SiO(2--1)} contour levels goes from \mbox{2 K km s$^{-1}$} in steps of \mbox{1 K km s$^{-1}$}.
The white crosses indicate the centers of the \mbox{30$'' \times$ 30$''$} squares where the shown spectra 
of \mbox{SiO} (black lines) and \mbox{HNCO} (thick blue lines) have been obtained.
The beam size of the \mbox{SiO(2--1)} and \mbox{HCN(4--3)} emissions are shown as open and filled circles,
respectively, in the bottom-left corner of the map. 
\label{fig5}}  
\end{figure*} 
%
%_____________________________________________________________________________________________________________________________
%
In this section we analyze the morphology and kinematics of the emission for every observed 
molecular species.
\mbox{Figure \ref{fig2}} shows the integrated intensity maps of the observed molecular lines in 9 
velocity intervals, from \mbox{$-$125} to \mbox{130 km s$^{-1}$}, selected to highlight the different 
components seen in the data cubes (velocity-channel maps are available in the online version
\mbox{Figs. \ref{appfig1}--\ref{appfig5}}).   
In this figure, we also show the \mbox{CS(1--0)} integrated intensity maps of \citet{Tsu99} for comparison.
The rms noise for the integrated-velocity maps ($\sigma_{\Delta\mathrm{v}}$) has been calculated from the 
rms noise of the original data cubes ($\sigma_\mathrm{ch}$; \mbox{Sect. \ref{Obs}}) using the formula
\begin{equation}
\label{eq1}
 \sigma_{\Delta\mathrm{v}} = \frac{\sigma_\mathrm{ch}}{\sqrt{N_\mathrm{ch}}} \cdot \Delta\mathrm{v} \label{che1}
\end{equation}
where $N_\mathrm{ch}$ is the number of integrated velocity channels and $\Delta{\rm v}$ the velocity 
range where the emission is integrated.\\
\indent \mbox{Figure \ref{fig3}} shows \mbox{declination-velocity} maps at constant right ascension. 
To increase the signal-to-noise ratio and to clearly show the gas kinematics of the components 
associated with the most outstanding features, we have also averaged the declination-velocity cuts at 
four selected right ascension intervals.\\
\indent Finally, in \mbox{Fig. \ref{fig4}} we compare the integrated \mbox{SiO(2--1)} emission maps 
with the radio continuum image at \mbox{6 cm} \citep{YZM87}, where we can distinguish the thermal emission from 
\mbox{Sgr A West} (the gray minispiral in \mbox{Fig. \ref{fig1}}).
This feature is composed of three ionized gas streamers:
the Western Arc (the southwestern photoionized inner edge of the CND; \citealp{SeL85}), and
the Northern and Eastern Arms (infalling ionized gas that has been stripped off of the CND or has
originated outside the CND; \citealt{LoC83,RoG93}).\\
%
%_____________________________________________________________________________________________________________________________
%
\subsection{The \mbox{SiO(2--1)} data cube.}
\label{DataCube_sio21} 
The \mbox{SiO(2--1)} emission appears above the \mbox{3$\sigma$} level in a wide velocity range: 
from \mbox{$-$125 km s$^{-1}$}, where the emission shows up in the southwest (relative to the
position of \mbox{Sgr A$^*$}) to \mbox{130 km s$^{-1}$}, where the emission disappears in the northeast. 
The \mbox{SiO(2--1)} morphology shows the kinematics of the rotating CND. The southern and northern lobes 
are located at the ends of the major axis with the maximum LOS velocities, and 
the gas with the lower radial velocities is located in the minor axis \citep{Mas95}.
The velocity pattern of the CND can be clearly seen in \mbox{Fig. \ref{fig3}} at the 
\mbox{RA= [$-$45$''$, 45$''$]} cut. 
The derived velocity gradient is \mbox{$\sim$2.2 km s$^{-1}$ arcsec$^{-1}$} (\mbox{$\sim$60 km s$^{-1}$ pc$^{-1}$}).
The southern lobe is stronger than the northern one, as in the case of all the previously \mbox{HCN} observations. 
\mbox{SiO} also shows the extension to the southwest detected by \citet{Mas95} in \mbox{HCN(3--2)}, which is very likely the
southernmost part of the Western Streamer, as discussed later.
The inclination of the CND traced by the \mbox{SiO} emission agrees with previous observations, with the major axis 
aligned along the Galactic plane.\\
\indent \mbox{Figure \ref{fig5}} shows a composition of the velocity-integrated maps of the \mbox{SiO(2--1)} emission
at the extreme CND velocities, \mbox{[$-$50, $-$20] km s$^{-1}$} and \mbox{[70, 100] km s$^{-1}$}, and the 
\mbox{HCN(4--3)} interferometric map of the CND obtained by \citet{MCa09} smoothed to \mbox{15$''$} resolution. 
This clearly shows that the \mbox{SiO(2--1)} emission completely traces the CND.\\
\indent In the most extreme blueshifted velocity ranges (\mbox{[$-$125, $-$50] km s$^{-1}$}), the \mbox{SiO(2--1)} line 
presents maximum emission at three locations: \mbox{($-$15$''$, $-$30$''$)}, \mbox{(95$''$, 25$''$)}, and 
\mbox{($-$40$''$, $-$70$''$)}. 
The first peak emission correspond to the southwest lobe of the CND (labeled as \mbox{SW CND} in \mbox{Fig. \ref{fig4}}; 
see also \mbox{Fig. \ref{fig5}}).
The maximum at \mbox{(95$''$, 25$''$)} belongs to \mbox{Cloud A} (see \mbox{Sect. \ref{Dis_dist}}), a small molecular 
cloud that seems to be isolated from the surrounding molecular gas (see also the 
\mbox{declination-velocity} plot at \mbox{RA= [45$''$, 105$''$]} in \mbox{Fig. \ref{fig3}}). 
In the \mbox{[$-$95, $-$50] km s$^{-1}$} panel of \mbox{Fig. \ref{fig4}} we can appreciate that this cloud is 
located inside the eastern edge of the SNR \mbox{Sgr A East}.
The last maximum \mbox{($-$40$''$, $-$70$''$)}, which achieves its highest intensity in the velocity range 
\mbox{[$-$50, $-$20] km s$^{-1}$}, is also related to this SNR, as seems to border its 
southwestern region (labeled as \mbox{Western St. South} in \mbox{Fig. \ref{fig4}}). 
\mbox{HCO$^+$(1--0)} and \mbox{HCN(1--0)} emission has also been detected at this location by \citet{Wri01}. 
They attribute the emission to a continuation of the CND or to a high-velocity streamer. However, our data 
suggest that this emission could belong to the southernmost part of the Western Streamer. It appears in 
the \mbox{declination-velocity} maps (panel labeled \mbox{RA= [$-$105$''$, $-$45$''$]} in 
\mbox{Fig. \ref{fig3}}) as a coherent structure with a constant velocity gradient 
(\mbox{$\sim$0.6 km s$^{-1}$ arcsec$^{-1}=$ 17 km s$^{-1}$ pc$^{-1}$}).\\  
\indent Between \mbox{$-$50} and \mbox{10 km s$^{-1}$}, the gas seems to connect the 
southwestern and northeastern regions, surrounding the \mbox{Sgr A$^*$} position. 
However, this feature is very likely the result of the superposition of several previously identified 
molecular features. 
The emission closer to \mbox{Sgr A$^*$} in projection comes from the eastern region of 
the CND \citep{Jac93,Mas95}. 
In previous \mbox{HCN(1--0)} single-dish observations, 
this part of the CND was absent because \mbox{HCN} emission was affected by self-absorption 
\citep{Mas95}. 
Even with the single-dish resolution, \mbox{SiO} emission perfectly traces the eastern region, 
although emission here is weaker than in the rest of the CND. 
Finally, the emission in this velocity range located in the northeastern quadrant (\mbox{$\Delta\delta >$40$''$}) 
comes from the Northern Ridge and borders, in projection, the northern edge of the SNR \mbox{Sgr A} East. 
Emission from the western side (\mbox{$\Delta\alpha <-$40$''$}) corresponds to the Western Streamer, whereas emission 
from the south comes from the northern part of the \mbox{20 km s$^{-1}$} GMC, both features bordering the 
western and southwestern edges of this SNR (see \mbox{Fig. \ref{fig4}}).\\
\indent In the velocity range \mbox{[10, 40] km s$^{-1}$}, we can see the northern region of the \mbox{20 km s$^{-1}$} 
GMC (the Southern Streamer) and \mbox{50 km s$^{-1}$} GMCs at east, together with the northern half of the Western Streamer, 
which peaks at \mbox{($-$60$''$, 20$''$)}, west of the CND (labeled as \mbox{Western St. North} in \mbox{Fig. \ref{fig4}}).
At \mbox{$-$5 km s$^{-1}$}, the \mbox{20 km s$^{-1}$} GMC begins to appear at the south. 
As the velocity increases, the molecular emission moves to 
the east to form other well-known features: the Molecular Ridge, a connection between the two 
most massive clouds of the \mbox{Sgr A} complex, the \mbox{20 km s$^{-1}$} and  
\mbox{50 km s$^{-1}$} GMCs. 
These three molecular features surround a cavity where the \mbox{SiO(2--1)} emission
has low intensity between \mbox{20} to \mbox{35 km s$^{-1}$}. 
At \mbox{70 km s$^{-1}$}, the \mbox{50 km s$^{-1}$} GMC finally disappears.
The \mbox{50 km s$^{-1}$} GMC is located on the eastern border of 
the SNR \mbox{Sgr A East} (see \mbox{Fig. \ref{fig4}}), 
whereas the molecular gas at the Southern Streamer could be affected by the interaction of  
two expanding shells: \mbox{Sgr A East} and \mbox{G 359.92$-$0.09} \citep{CoH00}.\\
\indent In the velocity range \mbox{[40, 70] km s$^{-1}$}, we can see another small 
cloud peaking at \mbox{($-$10$''$, 20$''$)}, just with a small offset east from the northern half of 
the Western Streamer. 
If we compare with the \mbox{HCN(4--3)} map of \citet{MCa09}, we can conclude that this cloud could 
be part of the northwestern region of the CND (labeled as \mbox{NW CND} in \mbox{Fig. \ref{fig4}}). 
In the velocity range \mbox{[70, 100] km s$^{-1}$}, the emission from this cloud moves to the 
east (maximum at \mbox{( 40$''$, 35$''$)}) showing a very good correlation with the northeastern 
region of the CND. \\
\indent Finally, in the highest velocity range (\mbox{[100, 130] km s$^{-1}$}), the maximum 
of the emission is located at the north of the CND, at \mbox{(35$''$, 55$''$)}.
In summary, \mbox{SiO(2--1)} emission traces all previously identified molecular structures 
within the central \mbox{12 $\times$ 12 pc$^2$} region. \\
%
%_____________________________________________________________________________________________________________________________
%
\subsection{The \mbox{H$^{13}$CO$^+$(1--0)} data cube.}
\label{DataCube_h13coP10} 
The \mbox{H$^{13}$CO$^+$(1--0)} emission covers the same velocity range as the \mbox{SiO(2--1)} emission and
nicely follows its morphology, except for the lower intensity of the \mbox{H$^{13}$CO$^+$} emission and the 
absorption features that appear toward \mbox{Sgr A$^*$}.
Two absorption features are observed (\mbox{Fig. \ref{appfig2}}): One narrow feature centered at 
\mbox{$-$52 km s$^{-1}$} arising from the \mbox{3-kpc} spiral arm \citep{Lin95} and observed in emission in 
the \mbox{C$^{18}$O(1$-$0)} maps (see \mbox{Sect. \ref{DataCube_c18o10}}); and a wide feature 
arising from local gas with a likely contribution from gas within the nuclear region, since the 
\mbox{SiO(2--1)} emission also shows a decrease in its intensity across this velocity range.
%
%_____________________________________________________________________________________________________________________________
%
\subsection{The \mbox{HN$^{13}$C(1--0)} data cube.}
\label{DataCube_hn13c10} 
The \mbox{HN$^{13}$C(1--0)} emission is weaker than that of \mbox{H$^{13}$CO$^+$(1--0)} and \mbox{SiO(2--1)}. 
Most of its emission is concentrated in the GMCs and in the Molecular Ridge,
going from \mbox{$-$20} to \mbox{70 km s$^{-1}$}. 
There is also low-intensity emission coming from the Northern Ridge 
visible in the velocity range \mbox{[$-$20, 10] km s$^{-1}$};
however, the eastern border of the CND is not traced by \mbox{HN$^{13}$C}.
In contrast, its western side seems to be detected at the noise level in the velocity ranges 
between \mbox{[40, 100] km s$^{-1}$}.   
For the \mbox{$-$82 km s$^{-1}$} velocity-channel map (\mbox{Fig. \ref{appfig3}}), we can see very low-intensity 
emission coming from Cloud A, however, in the corresponding velocity-integrated map there is no trace 
of this cloud.
At the velocity ranges between \mbox{[10, 40] km s$^{-1}$} there is low intensity emission 
from the northern half of the Western Streamer, as well as from its southern half in the velocity ranges 
between \mbox{[$-$95, 10] km s$^{-1}$}.\\
%
%_____________________________________________________________________________________________________________________________
%
\subsection{The \mbox{HNCO(5--4)} data cube.}
\label{DataCube_hnco54} 
This molecular line is the one that shows emission in the narrowest velocity range, only
from \mbox{$-$30} to \mbox{85 km s$^{-1}$} in the velocity-channel maps (\mbox{Fig. \ref{appfig4}}).
The HNCO emission is very intense only in the \mbox{20} and \mbox{50 km s$^{-1}$} GMCs, and in 
the Molecular Ridge that connects both clouds. It is remarkable that \mbox{HNCO(5$-$4)} emission
is the strongest of all the emission lines we have mapped, including \mbox{C$^{18}$O}. 
Outside these dense GMCs, the \mbox{HNCO} emission quickly drops. 
Between \mbox{$-$25} and \mbox{$-$5 km s$^{-1}$}, we can appreciate the HNCO counterparts of the 
Northern Ridge and the \mbox{20 km s$^{-1}$} GMC. 
However, as the molecular gas approaches the central position, it disappears.
We do not detect the \mbox{HNCO} emission from Cloud A observed in other molecular lines 
at negative velocities. There is, however, a very low-level emission from the southern and northern parts of 
the Western Streamer in the velocity ranges between \mbox{[$-$95, 40] km s$^{-1}$}.
Finally, the western side of the CND seems to be detected in the velocity range 
\mbox{[70, 100] km s$^{-1}$}.\\
%
%_____________________________________________________________________________________________________________________________
%
\subsection{The \mbox{C$^{18}$O(1--0)} data cube.}
\label{DataCube_c18o10} 
The \mbox{C$^{18}$O(1--0)} emission ranges from \mbox{$-$80} to \mbox{95 km s$^{-1}$} in the 
velocity-channel maps (\mbox{Fig. \ref{appfig5}}).
However, only for velocities \mbox{$\geq-$30 km s$^{-1}$} its emission seems to resemble 
the morphology observed in other molecular lines. We also find very low-intensity emission at around 
\mbox{170 km s$^{-1}$} detectable in the spatially-averaged spectrum over the whole region, probably  
arising from the expanding molecular ring \citep{Lin95}, which also appears in the \mbox{CS(1--0)} 
averaged spectrum.
Between \mbox{$-$80} to \mbox{$-$55 km s$^{-1}$}, the emission comes from the north, 
with no detected counterparts in the other molecules. 
Most of the narrow emission only seen in the \mbox{$-$52 km s$^{-1}$} channel map is foreground, 
arising from the \mbox{3-kpc} spiral arm \citep{Lin95}. This feature is observed in absorption
in the \mbox{H$^{13}$CO$^+$(1--0)} maps (see \mbox{Sect. \ref{DataCube_h13coP10}}). 
The Northern Ridge, as well as gas surrounding \mbox{Sgr A$^*$} with stronger emission at its 
western side, can be detected between \mbox{$-$20} and \mbox{10 km s$^{-1}$}. If we compare 
emission at these velocities with the \mbox{SiO(2--1)}, we notice that the \mbox{C$^{18}$O(1--0)} 
emission is shifted toward the west.
There is strong emission arising from the \mbox{20} and \mbox{50 km s$^{-1}$} GMCs,
and the Molecular Ridge, but we cannot clearly distinguish the emission from the northern part of 
the Western Streamer and Cloud A.
At the highest velocities, there is \mbox{C$^{18}$O(1--0)} emission coming from the north, probably 
associated with the northwestern lobe of the CND, but more extended to the northwest than the 
\mbox{SiO(2--1)} and \mbox{H$^{13}$CO$^+$(1--0)} emission at these velocities. 
The same happens to the \mbox{CS(1--0)} emission.\\
\indent From 100 to \mbox{130 km s$^{-1}$}, we find low-intensity emission
that resembles the morphology of the Molecular Ridge and the two GMCs 
of the \mbox{Sgr A} complex. 
This emission could come from \mbox{SO$_2$(17$_{5,13}$--18$_{4,14}$)} \citep{Lin95} and/or 
\mbox{NH$_2$CHO(5$_{1,4}$--4$_{1,3}$)} \citep{JAB84,Cum86}. \citet{Cum86} found this line in 
their spectral survey of \mbox{Sgr B2} and considered that the \mbox{SO$_2$} line contributed 
with 10$\%$. We also consider that the \mbox{NH$_2$CHO} line is the strongest one in this case.
The upper-level energy of the \mbox{NH$_2$CHO} transition is lower than the upper-level energy of
the \mbox{SO$_2$} transition (\mbox{19 K} vs. \mbox{202 K}). 
Moreover, the central velocity of the \mbox{NH$_2$CHO} line at the maximum of its emission is 
\mbox{45 km s$^{-1}$}, whereas the central velocity of the \mbox{SO$_2$} line is slightly redshifted 
(\mbox{56 km s$^{-1}$}) with respect to the lines from the other species (\mbox{$\simeq$49 km s$^{-1}$} 
on average). 
The origin of the \mbox{NH$_2$CHO} molecule could be related to the \mbox{HNCO} origin. \citet{Bis07} find
a strong correlation between both species in their survey of molecules suspected to be produced 
by the grain surface chemistry at several selected high-mass young stellar objects. This correlation 
suggests that \mbox{HNCO} and \mbox{NH$_2$CHO} molecules share a similar formation mechanism, i.e. grain chemistry. 
The \mbox{NH$_2$CHO} emission appears to be stronger in the \mbox{50 km s$^{-1}$} 
GMC, a molecular cloud affected by shocks driven by the SNR \mbox{Sgr A East}, that could eject this 
molecule to the gas phase.
In the following, we do not discuss this emission since it is rather weak, and it is only 
significantly detected toward the \mbox{50 km s$^{-1}$} GMC.\\
%
%_____________________________________________________________________________________________________________________________
%
\section{Physical conditions of the gas and column density estimates.}
\label{PhysConds}
%
%_____________________________________________________________________________________________________________________________
%
\begin{figure*}
\sidecaption
\includegraphics[width=14cm]{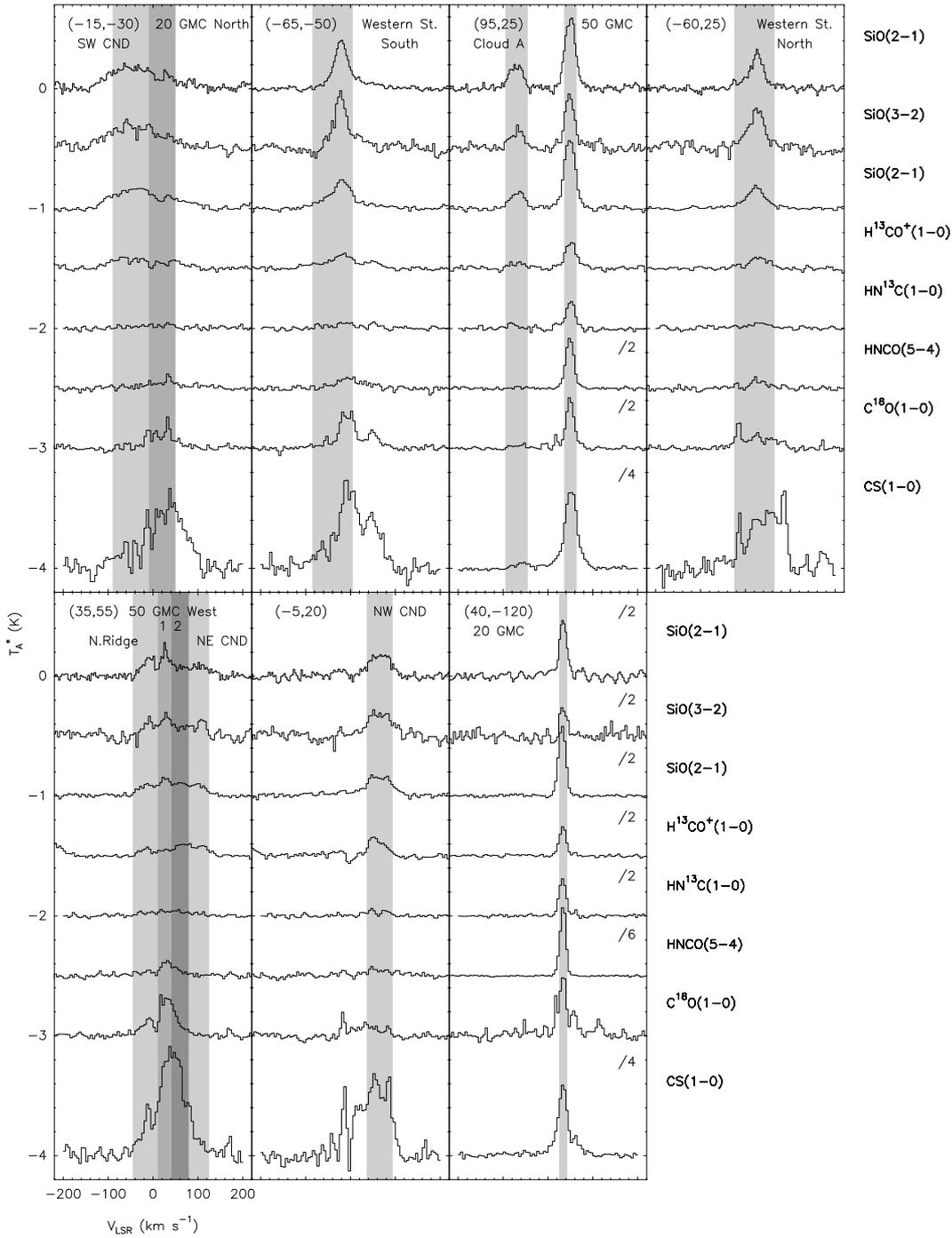}
\caption{Spectra of the observed transitions taken at 7 selected positions (marked with open squares in
\mbox{Fig. \ref{fig4}}).  
The upper \mbox{SiO(2--1)} and the \mbox{SiO(3--2)} spectra were observed simultaneously at the offset positions 
respect to \mbox{Sgr A$^*$} (in arcseconds) shown in the upper-left corner of each panel. 
The other \mbox{SiO(2--1)} spectra, together with the other molecular line 
spectra, were extracted from their respective data cubes from squares of area \mbox{30$'' \times$ 30$''$}.
Shadowed areas represent the selected velocity ranges where integrated intensities 
of the observed molecular lines were obtained, labeled with the names of their associated features.
The intensity of some transitions has been divided by a factor shown in the upper-right corner of 
the corresponding spectrum. 
\label{fig6}}  
\end{figure*}
%
%_____________________________________________________________________________________________________________________________
%
\begin{landscape}
\begin{table}
\begin{minipage}[t]{17cm}
\caption{Integrated intensities in selected velocity ranges. 
\label{tab2}}
\centering
\renewcommand{\footnoterule}{}  % to avoid a line before footnotes
\begin{tabular}{lllcccccccc}
\hline\hline
 Offset ($"$) & Component & $\Delta$v (km s$^{-1}$) &  \multicolumn{6}{c}{$\int T_{\rm A}^*$dv (K km s$^{-1}$)} \\ 
               &           &                         &  SiO(2$-$1) & H$^{13}$CO$^+$(1$-$0) & HN$^{13}$C(1$-$0) &  HNCO(5$-$4) & CS(1$-$0) &  C$^{18}$O(1$-$0)\\
\hline
($-$15, $-$30)  & SW CND                & [$-$90,$-$10] & 11.79$\pm$0.24 & 5.91$\pm$0.24 & 1.19$\pm$0.24 &  1.2$\pm$0.3    &   12$\pm$1    &  2.4$\pm$0.3   \\
                & 20 GMC North          & [$-$10, 50]   &  5.47$\pm$0.21 & 2.49$\pm$0.21 & 1.62$\pm$0.21 & 3.26$\pm$0.23   & 26.1$\pm$0.8  & 7.15$\pm$0.23  \\
($-$65, $-$50)  & Western St. South     & [$-$85, 5]    & 12.32$\pm$0.23 & 6.70$\pm$0.23 & 2.61$\pm$0.23 &  3.6$\pm$0.3    & 28.8$\pm$1.0  & 10.9$\pm$0.3   \\
(95, 25)        & Cloud A               & [$-$95, $-$45]&  4.99$\pm$0.22 & 1.78$\pm$0.22 & 0.95$\pm$0.22 &     $\leq$ 0.8  &  5.3$\pm$0.6  &   2.0$\pm$0.3  \\
                & 50 GMC                & [36, 64]      & 13.48$\pm$0.17 & 5.04$\pm$0.17 & 4.71$\pm$0.17 &16.73$\pm$0.20   & 61.9$\pm$0.5  & 16.16$\pm$0.20 \\
($-$60, 25)     & Western St. North     & [$-$25, 65]   &  8.79$\pm$0.23 & 4.82$\pm$0.23 & 2.48$\pm$0.23 &  3.1$\pm$0.3    & 30.6$\pm$1.0  &   8.3$\pm$0.3  \\
 (35, 55)       & Northern Ridge        & [$-$45, 10]   &  4.12$\pm$0.18 & 2.15$\pm$0.18 & 1.28$\pm$0.18 & 1.55$\pm$0.25   & 13.1$\pm$0.7  &  4.40$\pm$0.25 \\
                & 50 GMC West I	        & [10, 40]	&  3.81$\pm$0.14 & 1.44$\pm$0.14 & 1.14$\pm$0.14 & 2.86$\pm$0.18   & 21.3$\pm$0.5  &  8.20$\pm$0.18 \\
                & 50 GMC West II        & [40, 80]	&  4.03$\pm$0.16 & 3.33$\pm$0.16 & 1.45$\pm$0.16 & 1.83$\pm$0.21   & 26.7$\pm$0.6  &  4.93$\pm$0.21 \\
                & NE CND                & [80, 125]     &  3.60$\pm$0.17 & 3.44$\pm$0.17 & 0.75$\pm$0.17 &     $\leq$ 0.7  &  7.0$\pm$0.6  &      $\leq$ 0.7\\
($-$5, 20)      & NW CND                & [36, 94]      &  8.38$\pm$0.20 & 6.02$\pm$0.20 & 1.85$\pm$0.20 &  2.5$\pm$0.3    & 31.5$\pm$0.6  &   3.2$\pm$0.3  \\
(40, $-$120)    & 20 GMC                & [25, 43]      & 17.07$\pm$0.15 & 7.06$\pm$0.15 & 9.06$\pm$0.15 &47.84$\pm$0.24   & 36.8$\pm$0.5  &  7.32$\pm$0.24 \\
\hline
\end{tabular}
\footnotetext{Note.--- When molecular lines are undetected ($<$3$\sigma$), we use the upper limit to the integrated intensity 
derived from the spectrum rms noise.\\}
\end{minipage}
\end{table}
\end{landscape}
%
%_____________________________________________________________________________________________________________________________
%
\begin{landscape}
\begin{table}
\begin{minipage}[t]{17cm}
\caption{Derived physical conditions of the gas and column densities in selected positions. 
\label{tab3}}
\centering
\renewcommand{\footnoterule}{}  % to avoid a line before footnotes
\begin{tabular}{lllccccccccc}
\hline\hline
 Offset & Component & $\Delta$v & $\tau_{\rm SiO}$ & $T_{\rm ex}^{\rm SiO}$ & $n_{\rm H_2}$ & $N_{\rm SiO}$ & $N_{\rm H^{13}CO^+}$ &  $N_{\rm HN^{13}C}$ & $N_{\rm HNCO}$ & $N_{\rm CS}$ & $N_{\rm C^{18}O}$ \\
 ($"$)  &           &(km s$^{-1}$) &  & (K) & ($\times$ 10$^5$ cm$^{-3}$) & \multicolumn{5}{c}{($\times$ 10$^{13}$ cm$^{-2}$)} & ($\times$ 10$^{16}$ cm$^{-2}$) \\
\hline
($-$15, $-$30)     & SW CND                &   [$-$90,$-$10]   & 0.03   &  9   & 3    &  1.9  & 0.5    & 0.17  & 1.4        &  15  &  0.6        \\
                   & 20 GMC North          &   [$-$10, 50]     & 0.020  &  9   & 4    &  0.8  & 0.25   & 0.23  & 4          &  30  &  1.9        \\
($-$65, $-$50)     & Western St. South     &   [$-$85, 5]      & 0.03   &  8   & 3    &  1.9  & 0.6    & 0.4   & 4          &  30  &  3          \\
(95, 25)           & Cloud A               &   [$-$95, $-$45]  & 0.03   &  6   & 1.8  &  0.8  & 0.13   & 0.15  & $\leq$ 0.8 &   5  &  0.5        \\
                   & 50 GMC                &   [36, 64]        & 0.15   &  6   & 1.5  &  2.4  & 0.4    & 0.8   & 16         &  60  &  4          \\

($-$60, 25)        & Western St. North     &   [$-$25, 65]     & 0.017  & 12   & 5    &  1.4  & 0.5    & 0.4   & 4          &  50  &  2.2        \\
(35, 55)           & Northern Ridge        &   [$-$45, 10]     & 0.022  &  6   & 1.5  &  0.7  & 0.16   & 0.22  & 1.5        &  11  &  1.1        \\
                   & 50 GMC West I         &   [10, 40]        & 0.04   &  7   & 2.4  &  0.6  & 0.12   & 0.17  & 3          &  22  &  2.1        \\
                   & 50 GMC West II        &   [40, 80]        & 0.014  &  8   & 3    &  0.6  & 0.3    & 0.21  & 2.1        &  30  &  0.9        \\ 
                   & NE CND                &   [80, 125]       & 0.012  & 10   & 5    &  0.6  & 0.4    & 0.12  & $\leq$ 0.8 &  10  &  $\leq$0.17 \\ 
($-$5, 20)         & NW CND                &   [36, 94]        & 0.03   & 10   & 4    &  1.4  & 0.6    & 0.3   & 3          &  40  &  0.8        \\ 
(40, $-$120)       & 20 GMC                &   [25, 43]        & 0.5    & 4    & 0.4  &  6    & 0.6    & 3     & 50         &  25  &  1.5        \\  
\hline
\end{tabular}
\end{minipage}
\end{table}
\end{landscape}
%
%_____________________________________________________________________________________________________________________________
%
To estimate the column densities of the observed molecules, we selected several 
positions located at maxima of the \mbox{SiO(2--1)} emission. 
The spectra toward these positions can be seen in \mbox{Fig. \ref{fig6}}. 
These spectra show at least 9 different velocity components that can be fitted to Gaussian profiles, 
however, other components cannot be fitted. Moreover, in some cases it is not possible
to separate contributions from the different molecular features that appear at the same position
but in different velocity ranges using Gaussian profiles.
Therefore, we derived the column densities of the observed molecules from integrated intensities 
in selected velocity ranges (see \mbox{Table \ref{tab2}}). 
These selected velocity ranges are shown as shadowed areas in \mbox{Fig. \ref{fig6}}, labeled with their 
associated molecular feature, and the location of most of these components in the 
\mbox{$\alpha$-$\delta$-v} space is indicated in \mbox{Fig. \ref{fig4}}.
We note that the components \mbox{20 GMC North} and \mbox{50 GMC West} are very likely contaminated
with molecular gas from the CND.\\
\indent The procedure used to derive the column densities is the following. We constrained combinations of the pair
kinetic temperature (\mbox{$T_{\rm k}$}) and volume density (\mbox{$n_{\rm H_2}$}), together with the \mbox{SiO} 
column density (\mbox{$N_{\rm SiO}$}), by matching the two observed transitions of \mbox{SiO} (two upper spectra 
in each panel of \mbox{Fig. \ref{fig6}}) using a large velocity gradient (LVG) excitation code.
We explored a range of \mbox{$T_{\rm k}$} from 30 to \mbox{200 K}, as derived from 
ammonia inversion lines by \citet{Hut93} in several clouds of the GC and by \citet{HeH05} in the central 
\mbox{10--20 pc} region. 
Both studies conclude that molecular clouds in the GC can be roughly characterized by a 
two-temperature gas distribution in which 75$\%$ of the column density comes from cool gas (\mbox{$\sim$25 K}) 
and the remaining 25$\%$ from hot gas (\mbox{$\sim$200 K}). 
Furthermore, \citet{RFe01} find, from observations of \mbox{S(0)} to \mbox{S(5)} H$_2$ pure-rotational lines toward
several molecular clouds of the GC, that warm gas (\mbox{$\sim$150 K}) represents the \mbox{$\sim$30$\%$}
of the gas traced by low-J \mbox{CO} isotopes.
Then, we modeled the intensity of the other molecular lines (of \mbox{HNCO(5--4)} and 
\mbox{C$^{18}$O(1--0)} lines using the RADEX LVG excitation code; \citealp{Tak07}) with the previously 
derived physical condition pairs (\mbox{$T_{\rm k}$} and \mbox{$n_{\rm H_2}$}), in order to obtain their
column densities.
As all the observed transitions, except \mbox{C$^{18}$O(1--0)}, have similar dipole moments 
(from 1.60 to 3.89 D), we can assume that all transitions sample the same molecular gas 
(i.e., gas with similar physical conditions). 
However, for the \mbox{C$^{18}$O(1--0)} line (0.11 D), the derived column density values should be 
taken with caution.\\
\indent In \mbox{Table \ref{tab3}} we present the results of this analysis for a \mbox{$T_{\rm k}$} of \mbox{50 K}.
Raising the kinetic temperature by a factor of 4 has almost no effect on the excitation temperature and column 
density of \mbox{SiO}, but it decreases the volume density by a factor of \mbox{$\sim$3}, except for the 
\mbox{20 km s$^{-1}$} GMC, where the volume density decreases by a factor of 4.
The derived \mbox{$n_{\rm H_2}$} values at \mbox{$T_{\rm k}$ =50 K}
are very similar for all the gas components, \mbox{(1.5--5) $\times$ 10$^5$ cm$^{-3}$}, 
except for the \mbox{20 km s$^{-1}$} GMC, where the volume density shows the minimum value 
(\mbox{$\sim$4 $\times$ 10$^4$ cm$^{-3}$}). 
The \mbox{SiO} emission is optically thin, showing maximun optical depths of 0.5 toward the 
\mbox{20} and \mbox{$\sim$0.15} toward \mbox{50 km s$^{-1}$} GMCs. The \mbox{SiO} transitions are subthermally excited,
with \mbox{$T_{\rm ex}$} ranging from \mbox{4 K} at the \mbox{20 km s$^{-1}$} GMC to \mbox{12 K} in the northern 
part of the Western Streamer, in agreement with previous estimates in the GC \citep{Hut98}.
The \mbox{SiO} column densities, \mbox{0.6--6 $\times$ 10$^{13}$ cm$^{-2}$}, show less variation than the 
\mbox{HNCO} column densities, \mbox{$\leq$0.8--50 $\times$ 10$^{13}$ cm$^{-2}$}.\\
\indent HNCO column density presents the strongest contrast, with a difference of a factor of \mbox{$\geq$60} 
between Cloud A and the \mbox{20 km s$^{-1}$} GMC.
\mbox{HNCO} excitation temperatures show larger variations, depending on the volume density. 
Except for the \mbox{20 km s$^{-1}$} GMC, where the excitation temperature ranges between 9 and \mbox{15 K}, 
the HNCO excitation temperatures are in general \mbox{$\geq$40 K}. 
These values agree with the rotational temperatures derived by \citet{Mar08} using the rotational 
diagrams obtained from several \mbox{HNCO} transitions. 
They obtained \mbox{$T_{\rm rot}$ =35$\pm$12 K} and \mbox{$T_{\rm rot}$ =13.6$\pm$0.2 K} in their positions 
($-$30$''$, $-$30$''$) and ($-$15$''$, $-$215$''$) relative to \mbox{Sgr A$^*$}, respectively.
The first position almost coincides with our ($-$15$''$, $-$30$''$) position, whereas the second is located 
at the core of the \mbox{50 km s$^{-1}$} GMC. 
Therefore, our results share a similar trend, that is, the lowest HNCO excitation temperatures are shown toward 
the cores of the GMCs.
\mbox{HNCO} column densities show almost no variation for the different \mbox{$T_{\rm k}$} and 
\mbox{$n_{\rm H_2}$} pair values derived from the \mbox{SiO} transitions. The largest variation is found for 
the \mbox{20 km s$^{-1}$} GMC position with only a factor 1.6 more at \mbox{200 K} than at \mbox{50 K}. 
The \mbox{HNCO} emission is optically thin in all the cases (\mbox{$\leq$0.2}).\\
\indent In general, uncertainties in the column densities due to calibration (assuming a conservative 20$\%$),
and rms noise in the spectra are lower than a factor of 2. 
As for HNCO and SiO, changing the kinetic temperature from 50 to \mbox{200 K} has almost no effect
on the \mbox{H$^{13}$CO$^+$}, \mbox{HN$^{13}$C}, and CS column densities. The largest variation is also found
for the \mbox{20 km s$^{-1}$} GMC position, with a factor of 1.2 and around 1.7 larger at \mbox{200 K} than 
at \mbox{50 K} for CS and the other species. 
The \mbox{C$^{18}$O} column densities present larger differences with kinetic temperature, with factors up to 3. 
This higher sensitivity to volume density is expected, as \mbox{C$^{18}$O} is the molecule with the lowest critical 
density among the species studied in this paper. 
Therefore, we conclude that our column density estimates for SiO, \mbox{H$^{13}$CO$^+$}, \mbox{HN$^{13}$C}, HNCO, and CS
could present uncertainties of up to a factor of 2.\\
%
%_____________________________________________________________________________________________________________________________
%
\section{Integrated intensity maps and fractional abundances.}
\label{FracAbun}
%
%_____________________________________________________________________________________________________________________________
%
\begin{table*}
\begin{minipage}[t]{17cm}
\caption{Fractional abundances. 
\label{tab4}}
\centering
\renewcommand{\footnoterule}{}  % to avoid a line before footnotes
\begin{tabular}{llcccccc}
\hline\hline
 Offset & Component & $X_{\rm SiO}$ & $X_{\rm H^{13}CO^+}$ &  $X_{\rm HN^{13}C}$ & $X_{\rm HNCO}$ & $X_{\rm C^{18}O}$   \\
 ($"$) &           &  \multicolumn{4}{c}{($\times$10$^{-9}$)}                                     &  ($\times$10$^{-7}$)\\
\hline
($-$15, $-$30)     & SW CND                & 1.3  & 0.4   & 0.12 & 1.0        &  4   \\
                   & 20 GMC North          & 0.24 & 0.07  & 0.07 & 1.1        &  5   \\
($-$65, $-$50)     & Western St. South     & 0.6  & 0.18  & 0.11 & 1.2        &  8   \\
(95, 25)           & Cloud A               & 1.8  & 0.3   & 0.3  & $\leq$1.7  & 11   \\
                   & 50 GMC                & 0.4  & 0.06  & 0.14 & 3          &  7   \\
($-$60, 25)        & Western St. North     & 0.3  & 0.11  & 0.08 & 0.9        &  5   \\
(35, 55)           & Northern Ridge        & 0.6  & 0.15  & 0.20 & 1.4        & 10   \\
                   & 50 GMC West I         & 0.3  & 0.05  & 0.08 & 1.4        &  9   \\
                   & 50 GMC West II        & 0.20 & 0.09  & 0.07 & 0.7        &  3   \\ 
                   & NE CND                & 0.6  & 0.4   & 0.12 & $\leq$0.8  &  $\leq$1.8 \\ 
($-$5, 20)         & NW CND                & 0.3  & 0.14  & 0.06 & 0.7        &  1.9 \\ 
(40, $-$120)       & 20 GMC                & 2.4  & 0.25  & 1.3  & 19         &  6   \\  
%----------------------------------------------------------------------------------------------------------------------------------------
\hline
\end{tabular}
\footnotetext{Note.--- Fractional abundances with respect to \mbox{H$_2$}, derived from the CS column density assuming 
$X_{\rm CS}$=10$^{-8}$ for the GC.}
\end{minipage}
\end{table*}
%
%_____________________________________________________________________________________________________________________________
%
\begin{figure*}
\centering
\includegraphics[width=5cm,angle=-90]{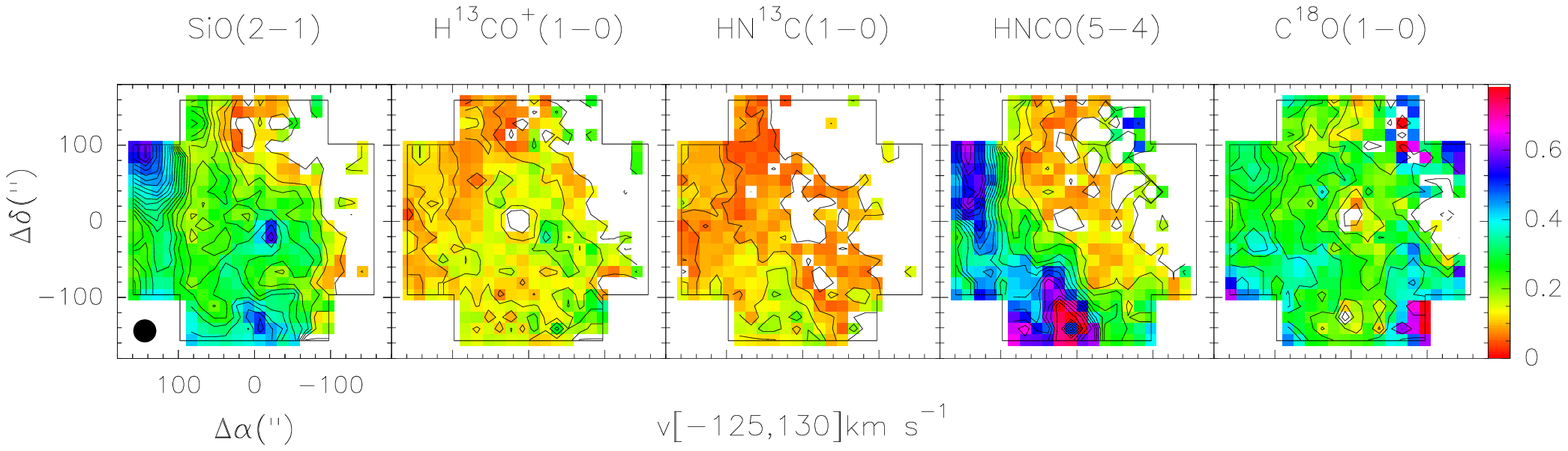}
\caption{\mbox{X/CS(1--0)} intensity ratios in the full velocity range shown at the bottom of the figure.
X stands for \mbox{SiO(2--1)}, \mbox{H$^{13}$CO$^+$(1--0)}, \mbox{HN$^{13}$C(1--0)}, \mbox{HNCO(5--4)}, and \mbox{C$^{18}$O(1--0)},
whose emission is shown in contour levels.
Intensity ratios have been derived considering only pixels with emission above the \mbox{3$\sigma$} level.
The wedge to the right shows the color scale of the intensity ratios. 
Contour levels are \mbox{$-$3$\sigma$} (dashed contour) and from \mbox{3$\sigma$} in steps of \mbox{4$\sigma$} 
(3.9 and \mbox{5.2 K km s$^{-1}$} for SiO/H$^{13}$CO$^+$/HN$^{13}$C and HNCO/C$^{18}$O ratios, respectively).
\label{fig7}}  
\end{figure*}
%
%_____________________________________________________________________________________________________________________________
%
\begin{figure*}
\centering
\includegraphics[width=8cm,angle=-90]{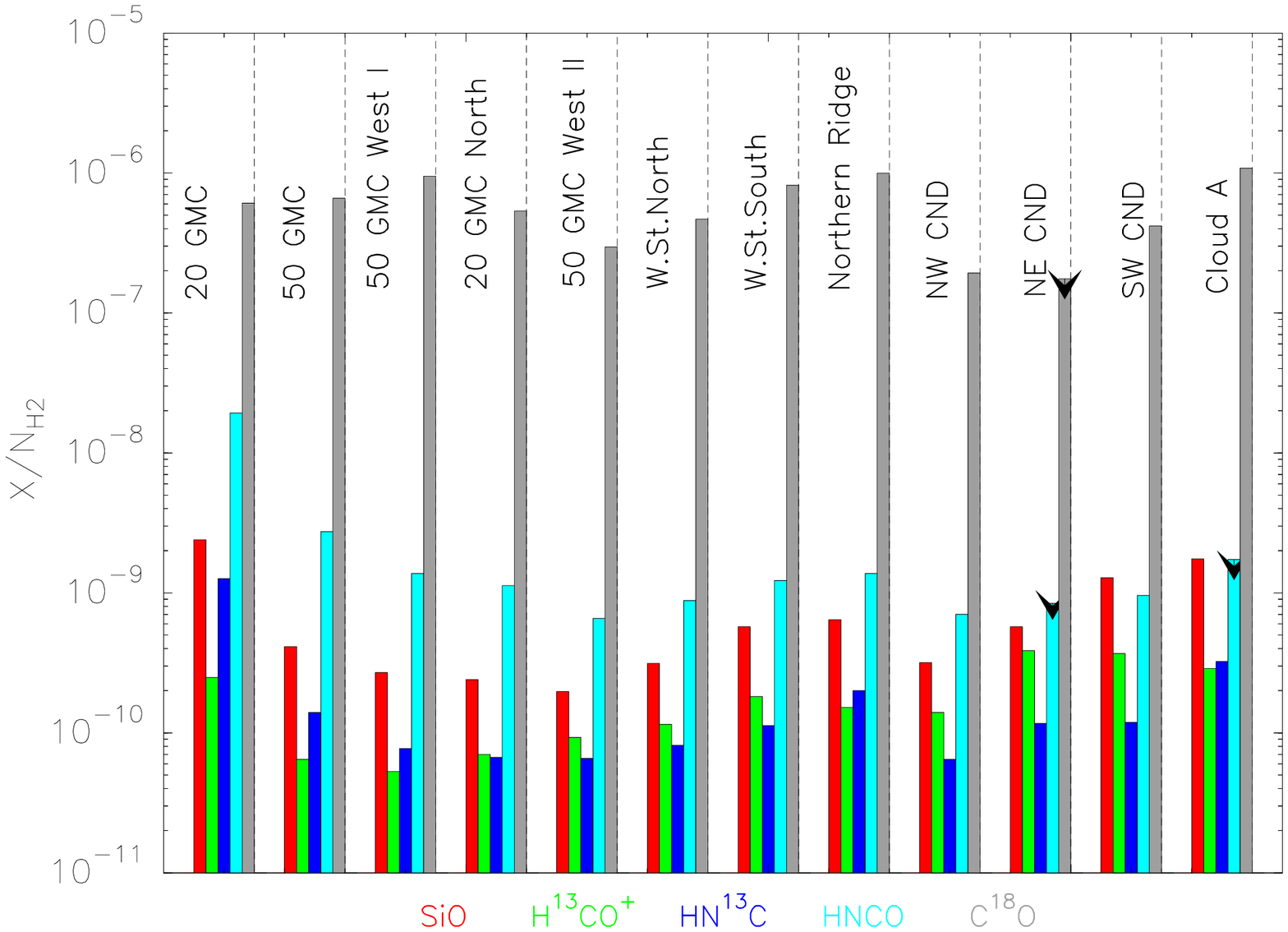}
\caption{Histogram with the fractional abundances shown in \mbox{Table \ref{tab4}}.
X stands for \mbox{SiO} (red), \mbox{H$^{13}$CO$^+$} (green), \mbox{HN$^{13}$C} (blue), \mbox{HNCO}
(cyan), and \mbox{C$^{18}$O} (gray), from left to right for every source.
\label{fig8}}  
\end{figure*}
%
%_____________________________________________________________________________________________________________________________
%
\begin{figure}
\centering
\includegraphics[width=9cm]{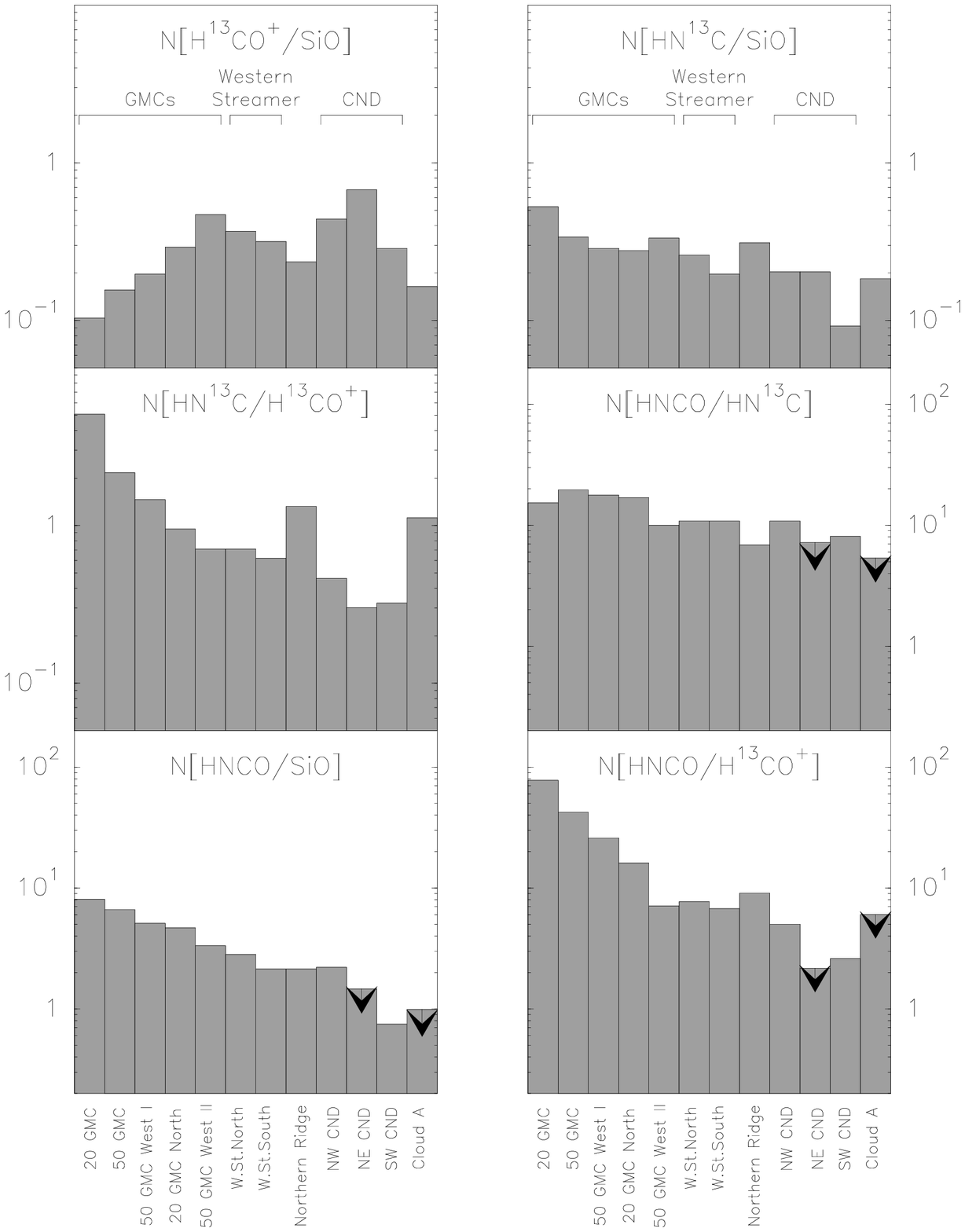}
\caption{Histogram with ratios between column densities shown in \mbox{Table \ref{tab3}}.
\label{fig9}}   
\end{figure}
%
%_____________________________________________________________________________________________________________________________
%
\indent To derive the fractional abundances of the observed species, we need to evaluate the hydrogen column density
at the selected positions. As mentioned before, \mbox{C$^{18}$O} is not the best tracer because it 
samples molecular gas biased toward low hydrogen volume densities, and it is strongly affected by gas along the LOS
toward the GC. 
Alternatively, CS is more likely arising from the same high-density molecular gas as
the other observed molecules, and it is moderately affected by any particular chemistry 
taking place at the central 12 parsecs.
It is only marginally enhanced in UV \citep{Goi06,Mar08} and shock-dominated environments \citep{Req06}.
However, the drawback is that \mbox{CS(1--0)} could be moderately affected by opacity.\\
\indent To elucidate the effects of CS opacity in our estimation of the fractional abundances,
we made integrated intensity ratio maps relative to the \mbox{CS(1--0)} emission (\mbox{Fig. \ref{fig7}}). 
Additionally, we present intensity ratio maps made in the same selected velocity ranges as \mbox{Fig. \ref{fig2}}
(\mbox{Fig. \ref{appfig6}}) in the online version. 
As we can see in \mbox{Figs. \ref{fig7}} and \mbox{\ref{appfig6}}, 
the \mbox{C$^{18}$O(1--0)/CS(1--0)} ratio is rather uniform.
The highest values are found at negative velocities, very likely associated with foreground molecular gas, 
and the lowest values at positive velocities associated with the northern part of the CND. 
These lower values indicate that the molecular gas is denser in the CND than in the surrounding GMCs 
(see \mbox{Table \ref{tab3}}). Therefore, \mbox{CS(1--0)} emission does not seem to be severely affected 
by optical depth effects.\\
\indent Using CS as a high-density molecular gas tracer and assuming a CS fractional abundance of 
\mbox{$X_{\rm CS}=N_{\rm CS}/N_{\rm H_2}$=10$^{-8}$} \citep{IGH87} for the GC, we derive hydrogen 
column densities between \mbox{0.5--6$\times$10$^{22}$ cm$^{-2}$}. 
As already mentioned, \mbox{Table \ref{tab4}} and \mbox{Fig. \ref{fig8}} show that \mbox{C$^{18}$O} is the molecule 
that presents the smallest variation in its fractional abundance, with only a difference of a factor of 6. 
On the other hand, HNCO is the molecule that presents
the largest variation of a factor of 28, with the highest values toward the cores of the GMCs, 
and the lowest ones toward the positions associated with the CND. 
Conversely, \mbox{SiO} presents a smaller variation of a factor of 12, and 
does not follow the trend observed in HNCO. SiO fractional abundance is high at the cores of the GMCs,
as well as in the positions of the CND.   
A similar behavior to HNCO follows \mbox{HN$^{13}$C} (a factor of 20). 
\mbox{HN$^{13}$C} seems to be anticorrelated with \mbox{H$^{13}$CO$^+$}, which shows the smallest variation
in its fractional abundance after \mbox{C$^{18}$O} (a factor of 7).\\
\indent The same trends can be seen in the integrated intensity ratio maps. 
The highest \mbox{SiO(2--1)/CS(1--0)} ratios and SiO fractional abundances are reached at the southwest lobe of the 
CND and Cloud A, only comparable to the ratio found at the GMCs. 
The gas components associated with the whole CND, the Western Streamer, and the Northern Ridge 
also present higher ratios than the surrounding gas (see \mbox{Fig. \ref{appfig6}} in the online version).
The \mbox{H$^{13}$CO$^+$(1--0)/CS(1--0)} ratio seems to follow similar behavior to the 
\mbox{SiO(2--1)/CS(1--0)} ratio, except for the gas components associated with the GMCs. 
The \mbox{HN$^{13}$C(1--0)/CS(1--0)} and \mbox{HNCO(5--4)/CS(1--0)} ratios shares a similar behavior, reaching
the maximum values of their ratios and abundances at the \mbox{20 km s$^{-1}$} GMC, and decreasing
toward the nuclear region.\\
\indent \mbox{Figure \ref{fig9}} presents histograms with the ratios between column densities of 
SiO, \mbox{H$^{13}$CO$^+$}, \mbox{HN$^{13}$C}, and HNCO toward selected positions. 
Sources have been ordered from left to right according to the \mbox{HNCO/SiO} column density 
ratio. There seems to be a relation between the \mbox{HNCO/SiO} ratio and the distance to the nuclear region.
All sources related with the GMCs are grouped with the highest ratios, the Western Streamer 
and Northern Ridge present intermediate values, whereas ratios toward the CND and Cloud A show the lowest values.
Similar trends present all the other ratios, except for the \mbox{H$^{13}$CO$^+$/SiO} column density ratio. 
The \mbox{HNCO/HN$^{13}$C} ratio is rather uniform with only a factor of 4 difference between the maximum 
and minimum values. 
Variations in the \mbox{HN$^{13}$C/SiO}, \mbox{H$^{13}$CO$^+$/SiO}, and \mbox{HNCO/SiO} ratios are within factors 6, 7, and 8,
respectively, whereas variations in the \mbox{HN$^{13}$C/H$^{13}$CO$^+$} and \mbox{HNCO/H$^{13}$CO$^+$} ratios are the greatest,
with values of 17 and 40, respectively.\\   
\indent In summary, we can group the observed species in those whose abundances decrease (HNCO and \mbox{HN$^{13}$C})
and those whose abundances are constant or enhanced (SiO and \mbox{H$^{13}$CO$^+$}) toward the nuclear region.\\
\section{Discussion.}
\label{Dis}
%
%_____________________________________________________________________________________________________________________________
%
\begin{landscape}
\begin{table}
\begin{minipage}[t]{23cm}
\caption{Fractional abundances. 
\label{tab5}}
\centering
\renewcommand{\footnoterule}{}  % to avoid a line before footnotes
\begin{tabular}{lclcccccccclcc}
\hline\hline
 Source  & & & $X_{\rm SiO}$ & $X_{\rm H^{13}CO^+}$ &  $X_{\rm HN^{13}C}$ & $X_{\rm HNCO}$ & $X_{\rm CS}$ & $X_{\rm H^{13}CN}$ & $N_{\rm H_2}$                & & Chemistry & & References \\
         & & & \multicolumn{6}{c}{($\times$10$^{-9}$)}                                                                         &($\times$10$^{22}$ cm$^{-2}$) & &           & &            \\
\hline      
 SW CND                     &                              &           & 1.3                      & 0.4                & 0.12               & 1.0         & 10                   & ...                 & 1.5   &                             & PDR$+$Shock                   &                             &\\ 
 20 GMC                     &                              &           & 2.4                      & 0.25               & 1.3                & 19          & 10                   & ...                 & 2.5   &                             & Shock                         &                             &\\ 
\hline 
                            &                              &           & \multicolumn{6}{c}{\textbf{Galactic sources}}                                                                                 &      &                              &                               &                             &\\
 \multirow{2}{*}{Orion Bar} & \multirow{2}{*}{{$\Big\{$}}  & (20$''$, $-$20$''$) & 0.03           & 0.05               & ... / 0.014        & $\leq$0.015 & 23                   & ... / 0.07          & 6.5  & \multirow{2}{*}{{$\Big\}$}}  & \multirow{2}{*}{PDR}          & \multirow{2}{*}{{$\Big\}$}} & \multirow{2}{*}{1, 2}\\  
                            &                              & (60$''$, $-$60$''$) & ...            & ... / 0.08         & ... / 0.08         & $\leq$0.3   & 5                    & ... / 0.04          & 0.3  &                              &                               &                             &\\
 IC443$-$G                  &                              &           & 0.8                      & $\leq$0.1 / 0.05   & ... / 0.015        & ...         & 3                    & ... / 0.12          & 1.2  &                              & Shock                         &                             & 3 \\ 
 Orion KL$-$Plateau         &                              &           & 22                       & ...                & ...                & 7           & 11                   & ...                 & 21   &                              & Shock                         & \multirow{2}{*}{{$\Big\}$}} & \multirow{2}{*}{4, 5, 6}\\
 Orion KL$-$Hot Core        &                              &           & 3                        & ...                & ...                & 14          & 50                   & ...                 & 42   &                              & Hot Core                      &                             &\\
 TMC$-$1                    &                              &           & $\leq$6$\times$10$^{-4}$ & 0.18               & 0.20               & 1.5         & 4                    & 0.17                &  1   &                              & Ion-molecule                  &                             & 7 \\ 
\hline
                            &                              &           & \multicolumn{6}{c}{\textbf{Extragalactic sources}}                                                                            &      &                              &                               &                             &\\
 NGC 253                    &                              &           & 0.13                     & 0.04               & 0.025              & 1.6         & 6                    & 0.13                & 6.7  &                              & SB                            &                             & 8 \\  
 NGC 4945                   &                              &           & $\leq$0.13               & 0.10               & 0.3                & 4           & 4                    & 0.20                & 6.4  &                              & AGN$+$SB ring                 &                             & 8, 9 \\  
 \multirow{3}{*}{M 82}      & \multirow{3}{*}{{$\Bigg\{$}} & Southwest & $\leq$0.12               & 0.12               & ...                & $\leq$9     & 2.3                  & ...                 & 3.8  & \multirow{3}{*}{{$\Bigg\}$}} & \multirow{3}{*}{Evolved SB}   & \multirow{5}{*}{{$\Bigg\}$}}& \multirow{5}{*}{8, 10, 11}\\ 
                            &                              & Center    & ...                      & ... / 0.06         & $\leq$0.3 / 0.04   & $\leq$9     & 7                    & $\leq$0.24 / 0.09   & 3.8  &                              &                               &                             &\\
                            &                              & Northeast & ...                      & ...                & ...                & $\leq$0.12  & 4                    & ...                 & 2.4  &                              &                               &                             &\\
 \multirow{2}{*}{IC 342}    & \multirow{2}{*}{{$\Big\{$}}  & Center    & ...                      & ... / 0.04         & ... /0.06          & 3           & 4                    & ... / 0.14          & 1.3  &                              & Normal$-$nucleus              &                             &\\ 
                            &                              & North     & $\leq$0.9                & 0.4                & ...                & 5           & 6                    & $\leq$1.1           & 1.1  &                              & Normal$-$arm                  &                             &\\
 \multirow{3}{*}{IC 342}    & \multirow{3}{*}{{$\Bigg\{$}} & A         & 0.16                     & 0.12               & ... / 0.08         & $\leq$4     & ... / (30--80)       & ... / 0.22          & 2.3  &                              & Normal$-$nucleus              & \multirow{3}{*}{{$\Bigg\}$}}& \multirow{3}{*}{12, 13, 14} \\ 
                            &                              & D         & 0.4                      & 0.06               & ... / 0.04         & 5           & ... / $\leq$(8--23)  & ... / 0.14          & 1.7  & \multirow{2}{*}{{$\Big\}$}}  & \multirow{2}{*}{Normal$-$arm} & &\\
 (interf.)                  &                              & D$'$      & ...                      & ...                & ... / 0.03         & 8           & ... / $\leq$(6--16)  & ...                 & 2.0  &                              &                               & &\\
\hline
\end{tabular}
\footnotetext{
References.---(1) \citet{Jan95}; (2) \citet{Sch01}; (3) \citet{DJP93}; (4) \citet{Nur09}; (5) \citet{Ter10a}; (6) Tercero et al., in prep.;
              (7) \citet{Nur07}; (8) \citet{Mar06}; (9) \citet{Wan04}; (10) \citet{Mar09}; (11) \citet{Bay09}; (12) \citet{Dow92}; 
              (13) \citet{MeT05}; (14) \citet{Use06}.\\
Notes.--- 
\textit{Galactic sources}: The second value of the \mbox{H$^{13}$CO$^+$}, \mbox{HN$^{13}$C}, and \mbox{H$^{13}$CN} abundances 
has been derived from the main isotope assuming a \mbox{$^{12}$C/$^{13}$C} isotopic ratio of 65 for the Orion Bar 
and the SNR IC443$-$G.
\textit{Extragalactic sources}: a source size of \mbox{20$''$} \citep{Mar06,Wan04} has been assumed in 
our column density estimations of \mbox{M 82} and \mbox{IC 342} from single-dish observations, as well as optically 
thin emission and the LTE approximation.  
For that purpose, integrated intensities have been taken from the compilation made by \citet{Mar06} plus additional 
data taken from \citet{Mar09} and \citet{Bay09}.
The second values of the \mbox{H$^{13}$CO$^+$}, \mbox{HN$^{13}$C}, and \mbox{H$^{13}$CN} abundances have been derived 
from the main isotopes assuming \mbox{$^{12}$C/$^{13}$C$\simeq$ 40} \citep{Hen98}.
Interferometric IC 342 abundances have been derived from the parameters of the line profiles taken from \citet{Use06}, 
\citet{MeT05}, and \citet{Dow92}, assuming a source size of \mbox{6$''$}, optically thin emission, and the LTE approximation.
CS abundance has been derived from \mbox{C$^{34}$S} assuming \mbox{$^{32}$S/$^{34}$S$\sim$ 8--23} \citep{MeT05}.\\
}
\end{minipage}
\end{table}
\end{landscape}
%
%_____________________________________________________________________________________________________________________________
%
\begin{table*}
\begin{minipage}[t]{17cm}
\caption{Abundance ratios compared with prototypical Galactic and extragalactic sources. 
\label{tab6}}
\centering
\renewcommand{\footnoterule}{}  % to avoid a line before footnotes
\begin{tabular}{lclccccc}
\hline\hline
 Source                     &                              &           & SiO/CS                     & H$^{13}$CO$^+$/CS         & HN$^{13}$C/CS                    & HNCO/CS                  & HN$^{13}$C/H$^{13}$CN  \\  
\hline 
 SW CND                     &                              &           & 0.13                       & 0.04                      & 0.012                            & 0.10                     & ...                    \\  
 20 GMC                     &                              &           & 0.24                       & 0.025                     & 0.13                             & 1.9                      & ...                    \\  
 \hline 
                            &                              &           & \multicolumn{5}{c}{\textbf{Galactic sources}} \\
 \multirow{2}{*}{Orion Bar} & \multirow{2}{*}{{$\Big\{$}}  & (20$''$, $-$20$''$) & 1.2$\times$10$^{-3}$ & 2.0$\times$10$^{-3}$  & ... / 6$\times$10$^{-4}$         & $\leq$7$\times$10$^{-4}$ & ... / 0.20             \\  
                            &                              & (60$''$, $-$60$''$) & ...                  & ... / 0.015           & ... / 0.015                      & $\leq$0.07               & ... / 1.9              \\  
 IC443$-$G                  &                              &           & 0.3                        & $\leq$0.03 / 0.015        & ... / 5$\times$10$^{-3}$         & ...                      & ... / 0.13             \\  
 Orion KL$-$Plateau         &                              &           & 2.0                        & ...                       & ...                              & 0.6                      & ...                    \\ 
 Orion KL$-$Hot Core        &                              &           & 0.05                       & ...                       & ...                              & 0.3                      & ...                    \\  
 TMC$-$1                    &                              &           & $\leq$1.5$\times$10$^{-4}$ & 0.04                      & 0.05                             & 0.4                      & 1.2                    \\  
 \hline
                            &                              &           & \multicolumn{5}{c}{\textbf{Extragalactic sources}} \\
 NGC 253                    &                              &           & 0.020                      & 6$\times$10$^{-3}$        & 4$\times$10$^{-3}$               & 0.25                     &  0.20                  \\  
 NGC 4945                   &                              &           & $\leq$0.03                 & 0.025                     & 0.08                             & 1.0                      &  1.6                   \\  
 \multirow{3}{*}{M 82}      & \multirow{3}{*}{{$\Bigg\{$}} & Southwest & $\leq$0.05                 & 0.05                      & ...                              & $\leq$4                  & ...                    \\  
                            &                              & Center    & ...                        & ... / 9$\times$10$^{-3}$  & $\leq$0.04 / 6$\times$10$^{-3}$  & $\leq$1.2                & ... / 0.4              \\  
                            &                              & Northeast & ...                        & ...                       & ...                              & $\leq$0.03               & ...                    \\ 
 \multirow{2}{*}{IC 342}    & \multirow{2}{*}{{$\Big\{$}}  & Center    & ...                        & ... / 0.010               & ... / 0.014                      & 0.8                      & ... / 0.4              \\  
                            &                              & North     & $\leq$0.16                 & 0.07                      & ...                              & 1.0                      & ...                    \\  
 \multirow{3}{*}{IC 342}    & \multirow{3}{*}{{$\Bigg\{$}} & A         & (2.3--7)$\times$10$^{-3}$  & (1.7--5)$\times$10$^{-3}$ & ... / (1.1--3)$\times$10$^{-3}$  & $\leq$0.14               & ... / 0.23             \\  
                            &                              & D         & $\geq$0.018                & $\geq$3$\times$10$^{-3}$  & ... / $\geq$1.9$\times$10$^{-3}$ & $\geq$0.25               & ... / 0.18             \\  
 (interf.)                  &                              & D$'$      & ...                        & ...                       & ... / $\geq$1.8$\times$10$^{-3}$ & $\geq$0.5                & ...                    \\  
\hline
\end{tabular}
\footnotetext{
\mbox{Notes.---}When two values for the abundance ratio are presented, the first one has been derived from the $^{13}$C isotopic
species and the second one from the main isotope assuming \mbox{$^{12}$C/$^{13}$C$\approx$65}
for IC443$-$G and the Orion Bar \citep{DJP93,Jan95}, and \mbox{$^{12}$C/$^{13}$C$\simeq$ 40} \citep{Hen98} for the 
extragalactic sources.
Interferometric column density ratios of \mbox{IC 342} have been derived from \mbox{C$^{34}$S} assuming a range of \mbox{$^{32}$S/$^{34}$S} 
isotopic ratios of 8--23 (see \citealt{MeT05}).\\}
\end{minipage}
\end{table*}
%
%_____________________________________________________________________________________________________________________________
%
\indent In this section, we discuss the most outstanding chemical differences found in our data
in the context of the two main mechanisms considered to be the main drivers of the chemistry and the heating of 
the central region: shocks and UV radiation (see \mbox{Sect. \ref{intro}}).  
To establish the dominant chemistry for the GC clouds from the observational point of view, 
we have selected several sources that are well-known prototypes of the different kinds of chemistry in the Milky 
Way and in nearby galaxies (see \mbox{Table \ref{tab5}}). 
A brief description of the physical conditions of these prototypical sources is given 
before the discussion.
To avoid uncertainties associated with the different ways in which the fractional abundances
were calculated for the prototypical sources, we do not compare fractional abundances, 
but rather their ratios (see \mbox{Table \ref{tab6}}).
From all the sources in the central 12 pc where we have derived abundances, we have selected 
the SW CND and 20 GMC as the most representative positions for the chemistry in this region.
We find a difference of one order of magnitude between the \mbox{HNCO/CS} and \mbox{HN$^{13}$C/CS} abundance
ratios toward both sources, whereas the \mbox{SiO/CS} and \mbox{H$^{13}$CO$^+$/CS} ratios show very 
similar values.\\
\indent Finally, once we had defined the dominant chemistry drivers for each particular species, we used 
the effects that UV radiation produces on the molecular gas to establish the location of the 
different molecular components relative to the Central cluster.\\
%
%_____________________________________________________________________________________________________________________________
%
\subsection{Prototypical sources.}
\label{Dis_GalSou} 
\indent As Galactic prototypical sources, we selected a cold dark cloud, a photo-dissociated region (PDR), 
molecular clouds interacting with an SNR and molecular outflows, and two hot cores.\\ 
\indent TMC-1 is one of the best-studied quiescent dark clouds, with temperatures around \mbox{10 K} and densities of a 
few \mbox{10$^4$ cm$^{-3}$}. Due to the lack of internal heating sources and shocks, dark clouds are the best 
places to study gas-phase chemistry driven by ion-neutral reactions.\\
\indent As PDR, we have selected the \mbox{OMC$-$1} ridge. The UV radiation is provided mostly by the 
Trapezium stars, which are located at \mbox{0.25 pc} from the PDR.
The mean kinetic temperature in the molecular layers of this PDR is around \mbox{85 K} with \mbox{90$\%$} 
of the molecular gas in a homogeneous layer of density of \mbox{10$^4$ cm s$^{-1}$} and the remaining
material concentrated in high-density clumps of \mbox{$\sim$10$^6$ cm s$^{-1}$} \citep{HJD95}.
The UV radiation flux in the Orion PDR seems to be quite similar to that of the inner edge of the CND, 
with \mbox{$G{\rm _0}\sim$10$^{5}$} (in units of the Habing field; \citealp{BHT90}).\\
\indent To test shock chemistry, we use clump G in the SNR \mbox{IC443}, where shocks occur 
nearly perpendicular to the LOS. Densities and temperatures of the shocked gas are 
\mbox{10$^5$--3$\times$10$^6$ cm$^{-3}$} and \mbox{80--200 K}, respectively \citep{DJP93}.
Unfortunately, HNCO has not been studied toward molecular clouds affected by SNRs. 
However, HNCO among other species, have been observed toward the closest high-mass star-forming region 
in our Galaxy, the Orion KL cloud \citep{Ter10a}. 
Single-dish line surveys find several spectral components within a telescope beam. 
One of these components, known as the \textit{plateau}, presents a chemistry dominated by shocks
driven by molecular outflows. The physical conditions of this component is characterized by
temperatures and densities around \mbox{100--150 K} and \mbox{$\sim$10$^5$ cm$^{-3}$}, respectively
(see the introduction of \citealp{Per07}).\\
\indent As prototypical hot cores, \mbox{Sgr B2(N)} and the hot core component of Orion KL were selected. 
The first one is the most prominent hot core of the \mbox{Sgr B2} GMC complex located at the GC. 
It presents densities as high as \mbox{10$^7$ cm$^{-3}$} and temperatures that go from \mbox{20 K} to up to 
\mbox{80 K} averaged over a region of \mbox{1 pc}, being even higher
(\mbox{150 K}) on smaller scales (see \citealp{Num00}). 
The Orion hot core is a warm star-forming region internally heated by one or more young massive protostars.
It contains very dense clumps (\mbox{$\sim$ 10$^7$ cm$^{-3}$}) and presents temperatures between 
\mbox{165--400 K} (see the introduction of \citealp{Per07}).\\ 
\indent On the other hand, as extragalactic prototypical sources, we selected three starburst nuclei
in different evolutionary stages (\mbox{NGC 253}, \mbox{M 82}, and \mbox{NGC 4945}), and a normal spiral galaxy more 
similar to the Milky Way (\mbox{IC 342}).  
When one compares GC abundances with those of extragalactic nuclei measured with single-dish telescopes, 
it has to be taken into account that single-dish measurements represent average values over regions of 
hundreds of parsecs that contains a large number of GMCs. At best, interferometer observations are able 
to provide average abundances over individual GMCs (\mbox{$\sim$ 50 pc}).\\  
%
%_____________________________________________________________________________________________________________________________
%
\subsection{SiO chemistry.}
\label{Dis_sio}
\indent SiO is one of the best tracers of energetic mechanical phenomena in the ISM, with fractional abundances 
that changes by 5 orders of magnitude in dark clouds \mbox{$\sim$10$^{-12}$} to regions associated with 
shocks \mbox{$\sim$10$^{-7}$} \citep{MPi92,JSe05}.
Indeed, the most outstanding difference between TMC-1 and the central 12-pc sources is the \mbox{SiO/CS} 
abundance ratio, since it is more than \mbox{2--3} orders of magnitude larger in the \mbox{SW CND} and the \mbox{20 GMC}
than in cold clouds.
The reason for this huge variation is the depletion of \mbox{SiO} onto grains in quiescent clouds. 
Only when energetic mechanisms destroy or erode dust grains does \mbox{SiO} appear with significant abundance
in the gas phase.\\ 
\indent The widespread SiO emission in the GC, with typical abundances of \mbox{$\sim$10$^{-9}$} \citep{MPi97}, 
has led to the conclusion that large-scale shocks must be at work in the GC.
However, there are other energetic mechanisms, apart from shocks, that might enhance SiO abundances. 
It has been proposed that high-energetic phenomena, like X-rays as observed by the \mbox{6.4 keV} Fe line, 
enhance the \mbox{SiO} abundance \citep{MPi00,AAB09} in the GC over the averaged value.
But, the 20 and \mbox{50 km s$^{-1}$} GMCs do not seem to present
an enhancement of the \mbox{6.4 keV} Fe line, which would indicate a recent interaction of molecular gas
with X-rays.\\
\indent Another energetic mechanism that could enhance SiO abundance in PDRs is the desorption of grain mantles by UV-photons, 
as \citet{Sch01} have shown for the Orion Bar and the S140 PDR.
 However, the \mbox{SiO/CS} abundance ratio in the Orion Bar is 2 orders of magnitude lower than toward 
the \mbox{SW CND} and the \mbox{20 GMC}. 
Thus, UV photo-desorption does not seem to play an important role in the enhancement of SiO in the central 12-pc of the GC.\\
\indent On the other hand, the \mbox{SiO/CS} abundance ratio in IC443$-$G is very 
similar to those found toward the \mbox{SW CND} and the \mbox{20 GMC} sources. 
However, the \textit{plateau} component of Orion KL presents a \mbox{SiO/CS} abundance ratio that is
one order of magnitude greater than those found at our sources. 
According to the C-shock models of \citet{JSe08}, our observed SiO abundances ($\sim$10$^{-9}$) can 
be accounted for by the sputtering of grain mantles in shocks with velocities between 20 and \mbox{30 km s$^{-1}$}
(assuming that a significant amount of silicon is locked in the mantles).  
At higher shock velocities, sputtering of grain cores begins and gives rise to SiO abundances much higher
($\sim$10$^{-7}$--10$^{-6}$) than the ones derived toward the 12 central parsecs of the Galaxy.\\
\indent Therefore, neither X-rays nor photo-desorption due to the UV field from the Central 
cluster can explain the SiO abundance in the central 12-pc sources, with C-shocks with velocities
between \mbox{20--30 km s$^{-1}$} the most likely origin of its abundance.\\
%
%_____________________________________________________________________________________________________________________________
%
\subsection{HNCO chemistry.}
\label{Dis_hnco}
\indent Isocyanic acid (HNCO) has been observed in a wide variety of sources, from dark clouds to hot cores 
(see the introduction of \citealt{Tid09} for a brief review) with abundances of
\mbox{$\sim$10$^{-9}-$10$^{-8}$}. 
In the GC, HNCO emission has also been found to be particularly strong and widespread
\citep{Lin95,KuS96,Dah97,Sat02,MiI06,Mar08}.\\
\indent The \mbox{HNCO/CS} abundance ratio in the (20$''$, $-$20$''$) position of the Orion Bar is \mbox{2--3} 
orders of magnitude lower than in the central 12-pc sources. Moreover, among all the sources 
listed in \mbox{Table \ref{tab5}}, the Orion Bar is where one can find the lowest HNCO abundances. 
The behavior of HNCO in PDRs is completely opposite that of \mbox{CS}.
\citet{Mar08} have studied the \mbox{HNCO} emission toward several positions in the GC, finding that the 
\mbox{HNCO} abundance decreases by up to a factor \mbox{22} for the sources affected 
by photodissociation, with maximum values of \mbox{2$\times$10$^{-8}$} toward the cores of GMCs
and minimum values of \mbox{1$\times$10$^{-9}$} toward the sources close to massive stellar clusters. 
Moreover, this difference factor increases up to \mbox{30} when the \mbox{HNCO/CS} abundance 
ratio is compared.
The HNCO abundance toward the SW CND is one order of magnitude lower than toward the 20 GMC, which is consistent with
the idea that HNCO is very easily photodissociated by UV radiation, as the SW CND is closer to the Central cluster 
and/or not as well-shielded as the 20 GMC.\\
\indent \citet{Mar09b} have modeled the HNCO abundance in PDRs as a function of the visual extinction with \mbox{$T_{\rm K}$= 50 K}, 
\mbox{$n_{\rm H_2}$= 10$^5$ cm$^{-3}$}, and a UV radiation field of \mbox{$G_{\rm 0}$= 5$\times$10$^3$} in Habing units, similar 
to the physical conditions of galactic nuclei. In their coupled dense core-PDR model, they find observable HNCO abundances only for
\mbox{$A_{\rm v}\geq$ 5} (i.e. \mbox{$N_{\rm H_2}\sim$ 5$\times$10$^{21}$ cm$^{-2}$}), concluding that molecular clouds affected 
by UV radiation with observable HNCO abundances must contain well shielded dense cores where HNCO molecules are protected 
against UV photons.
According to \citet{Mar09b}, this could be the reason the \mbox{HNCO/CS} ratio show a wide range 
of values toward extragalactic nuclei, from the upper limit of 0.03 in \mbox{M 82$-$NE} to the high values 
(0.8--1.0) found toward \mbox{IC 342} and \mbox{NGC 4945}, with the ratio in \mbox{NGC 253} in between 
(0.25; see \mbox{Table \ref{tab6}}). 
The low ratio measured toward the GMCs of \mbox{M 82} is the result of large amounts of gas being 
affected by UV radiation in its nuclear region. 
This is due to the lower column densities that \mbox{M 82} GMCs present in comparison with GMCs of other 
extragalactic nuclei, where higher \mbox{HNCO/CS} ratios are found \citep{Mar09b}.\\ 
\indent Therefore, our low measured value for the \mbox{HNCO/CS} abundance ratio toward the SW CND reflects the 
effect of UV radiation on the fragile HNCO molecules. Column densities measured in the 12 central parsecs are 
\mbox{$\sim$10$^{22}$ cm$^{-2}$}, with the lowest values found toward Cloud A and the NE CND, just the sources 
where no emission of HNCO has been detected.\\
\indent In contrast to the destruction mechanism of HNCO, the origin of this species is not yet
well understood.
Gas-phase chemistry alone is not able to reproduce the observed HNCO abundances, predicting 
abundances on the order of  \mbox{10$^{-10}$} \citep{Igl77}.
Chemistry in post-shocked gas driven by neutral-neutral reactions has been invoked to reproduce
HNCO abundances on the order of \mbox{10$^{-8}$}, as the high temperatures achieved by shocks could 
overcome the energy barriers of these reactions (for more details see \citealp{TTH99,Tid09,RFe10}).
However, this formation pathway presents two main drawbacks:
the short times that shocks can maintain the required high temperatures and cannot detect \mbox{O$_2$} in 
the \mbox{Sgr A} complex (\mbox{$X_{\rm O_2}$ $\leq$ 1.2$\times$10$^{-7}$}; \citealp{San08}), a key molecule  
in this formation path of HNCO.\\ 
\indent The alternative to HNCO production in gas phase is formation through grain-surface chemistry.
Hot core chemical models \citep{GWH08,Tid09}, as well as gas-grain chemistry models in cold clouds \citep{HaH93},
generate high abundances of HNCO in the grain mantles (\mbox{10$^{-7}$--10$^{-5}$}). 
There are also observational clues that point to a grain-surface origin for HNCO: the $``$XCN$"$ infrared (IR) 
absorption feature of interstellar ices that could correspond to the \mbox{OCN$^-$} anion \citep{HMG01}.
Another observational proof is the correlation between the HNCO and SiO-integrated and peak intensities 
found by \citet{ZHM00}, with SiO lines broader than HNCO lines, which suggests a common origin. 
Besides, HNCO abundances are also enhanced in high-velocity gas \citep{ZHM00}.
Moreover, the largest \mbox{SiO/CS} and \mbox{HNCO/CS} abundance ratios among our selected 
prototypical Galactic sources are found at the \textit{plateau} component of the \mbox{Orion KL} cloud,
a high-mass star-forming region affected by several outflows.\\
\indent However, after the formation of HNCO on the grain mantles, according to hot core chemical models, 
HNCO is destroyed by reactions with radicals originating more complex species, such as \mbox{NH$_2$CHO}.
Under this scenario, the observed HNCO abundances in the gas-phase would be the result of the destruction of these 
more complex molecules \citep{GWH08,Tid09}. 
Ejection of more complex molecules with the subsequent photodissociation and production of HNCO in gas-phase also seems 
unlikely in the GC sources, as we observed the highest HNCO abundances toward the more shielded cores of the GMCs
and the lowest ones toward regions with strong UV fields that should favour the destruction of the complex molecules. 
Moreover, observational results seem to contradict the idea of HNCO as a \textit{second generation} species.
\citet{Bis07} have measured a set of complex organic molecules toward a sample of hot cores. 
HNCO and \mbox{NH$_2$CHO} belong to the same group, that is, they are evaporated from grain mantles and are very 
likely \textit{first generation} species (i.e. directly released from the grains), as their abundances 
toward several sources are correlated. \\
\indent Therefore, HNCO could share the same origin as SiO, shocks, but HNCO seems to be more sensitive to UV radiation.
\mbox{IC 342}, a similar nearby galaxy to the Milky Way, has been mapped with interferometry technique in 
several species, revealing an spiral galaxy composed of several GMCs with chemistries driven by different mechanisms.
The \mbox{SiO/CS} and \mbox{HNCO/CS} ratios toward the GMC closer to the stellar cluster (\mbox{GMC A}) in \mbox{IC 342} are 
lower than those found toward the GMCs in the spiral arms (\mbox{GMCs D} and \mbox{D$'$}).
The same trend is followed by our \mbox{SW CND} and \mbox{20 GMC} sources, with the upper limit to the \mbox{HNCO/CS} 
abundance ratio at the \mbox{GMC A} of \mbox{IC 342} similar to our derived value at the \mbox{SW CND} source.
The \mbox{GMC A} could be suffering the direct impact of the UV radiation, stellar winds, and SN explosions originated 
in the central cluster \citep{Sch08}. Therefore, both sources seem to be affected by strong UV radiation and shocks.
In contrast, the GMCs in the spiral arms of \mbox{IC 342} show higher lower limits to the \mbox{SiO/CS} and 
\mbox{HNCO/CS} ratios, which could approach the \mbox{20 GMC} ratios. This is consistent with the idea that 
the GMCs in the spiral arms are affected by shocks, but not strongly by UV radiation.
\citet{MeT05} also mapped methanol (\mbox{CH$_3$OH}), finding abundance enhancements that can be produced by 
large-scale mild shocks due to orbital dynamics, as \mbox{CH$_3$OH} cannot be ejected to the gas by evaporation 
and the emission of both species is widespread.
Thus, interferometric observations of \mbox{IC 342} seem to support the idea of grain chemistry as the origin 
of HNCO.\\ 
%
%_____________________________________________________________________________________________________________________________
%
\subsection{H$^{13}$CO$^+$ chemistry.}
\label{Dis_h13coP}
\indent The \mbox{H$^{13}$CO$^+$/CS} abundance ratio toward the Orion PDR presents a difference of almost an 
order of magnitude between the offset positions \mbox{(20$''$, $-$20$''$)} and \mbox{(60$''$, $-$60$''$)} 
with respect to the ionization front (IF; see \mbox{Table \ref{tab6}}; \citealp{Jan95})
This difference is due to a larger increase in the CS abundance toward the closest position to the IF, which 
is \mbox{(20$''$, $-$20$''$)}, whereas the \mbox{HCO$^+$} abundance mostly remains constant.
PDR-models show that higher \mbox{HCO$^+$/CS} abundance ratios are expected toward the superficial layers 
of PDRs than toward cool cores \citep{StD95}.
The exception to this trend could be the most UV-exposed layer of PDRs. There, ion destruction is very 
efficient through dissociative recombination reactions, as electron densities reach their maximum values.
In the Horsehead edge-on PDR, the \mbox{HCO} emission traces the edge of the PDR, whereas the 
\mbox{H$^{13}$CO$^+$} emission appears more shifted to the cloud interior \citep{Goi09}.\\  
\indent Our \mbox{H$^{13}$CO$^+$/CS} abundance ratios are similar to that of the 
\mbox{(60$''$, $-$60$''$)} position of the Orion Bar, although the SW CND ratio is slightly larger (by factors of 2$-$2.6) 
and more similar to the value found at the dark cloud TMC-1, despite the strong UV field that impinges on the CND. 
This could be due to the different extension sampled by the telescope beam in the Jansen et al. survey (0.04 pc) 
and our work (1.2 pc). Our beam samples not only the PDR, but also a more shielded part of the cloud. Thus, the 
\mbox{H$^{13}$CO$^+$/CS} abundance ratio could reflect this mixture, approaching the value found in dark clouds.\\
\indent Conversely to the \mbox{HCO$^+$} chemistry in PDRs and dark clouds, there is no agreement with a possible 
\mbox{HCO$^+$} enhancement in shocks.  
Studies of molecular outflows \citep{Gar92,Gir99} have found enhancements of the \mbox{HCO$^+$} abundance in shocked gas.
\mbox{HCO$^+$} abundance enhancement of one order of magnitude have been detected in the interface between high-velocity 
outflows and the surrounding quiescent gas in low-mass star-forming regions \citep{Hog98}. 
This enrichment would be produced by the sputtering of \mbox{CO} and \mbox{H$_2$O} to gas phase from ice mantles, 
followed by photochemical processing by UV radiation, generating C$^+$ that would react 
with \mbox{H$_2$O} to enhance the \mbox{HCO$^+$} abundance \citep{Raw00,Vit02}.
However, there are also chemical models and observational results that report a decrease in the \mbox{HCO$^+$}
abundance in the postshock gas with respect to the preshock gas (see \citealp{DJP93}; \citealp{Gar98} and references 
therein).\\
\indent Our SW CND and 20 GMC sources present values of the \mbox{H$^{13}$CO$^+$/CS} abundance ratio in 
between those of TMC-1, clump G, and the \mbox{(60$''$, $-$60$''$)} position of the Orion Bar. 
The highest value of this ratio is the one of TMC-1, a prototype of quiescent cloud. 
This suggests that the ion abundance in our sources is not increased by shocks, despite the favorable 
conditions that this region presents, i.e., shocks plus UV radiation.\\ 
%
%_____________________________________________________________________________________________________________________________
%
\subsubsection{The GC ionization rate.}
\label{Dis_h13coP_IonRate} 
\indent \citet{Oka05} claim that the GC presents a higher ionization rate than anywhere else in the Galaxy 
(\mbox{$\zeta= 2-7\times10^{-15} {\rm s^{-1}}$}). 
\mbox{H$^{+}_{3}$} absorption lines toward selected stars reveal the presence of extensive hot and diffuse 
clouds with high-velocity dispersions in the inner \mbox{200 pc} of the Galaxy.
\citet{Oka05} found broad absorptions due to \mbox{H$^{+}_{3}$(3,3)} in the direction of the 
\mbox{50 km s$^{-1}$} GMC, consistent with its velocity range. However, \citet{GBW89}, from CO 
in absorption, prefer to establish the absorption origin at the inner edge of the CND.\\
\indent \citet{FMH94} have studied the effects of varying the cosmic-ray ionization rate in dark clouds. 
Their model predicts an increase in the \mbox{HCO$^+$} abundance as ionization rate becomes higher. 
Our \mbox{HCO$^+$} abundances, derived taking a \mbox{$^{12}$C/$^{13}$C} isotopic ratio 
of \mbox{20--25} \citep{Wil99} into account, are between \mbox{$\sim$10$^{-9}$--10$^{-8}$}. The maximum value corresponds to
the SW CND that, according to Farquhar et al.'s model, could be explained by the $``$standard$"$ cosmic-ray 
ionization rate (\mbox{$\zeta\sim$10$^{-17}$ s$^{-1}$}).
On the other hand, the minimum value, corresponding to the 50 GMC, could be produced by an 
ionization rate a factor of \mbox{$\sim$60} lower than the $``$standard$"$ value.
Therefore, a small gradient in the ionization rate could be responsible for the variations found in the 
\mbox{H$^{13}$CO$^+$} abundances toward the CND and the 50 GMC.
However, our values do not support the high ionization rate derived by \citet{Oka05}.  
When modeling of the observed intensities of \mbox{H$_2$O} and \mbox{H$_3$O$^+$} toward the \mbox{Sgr B2} envelope,  
\citet{vdT06} obtain a cosmic-ray ionization rate of \mbox{$\sim$ 4$\times$10$^{-16}$ s$^{-1}$}, 
lower than the value derived by \citet{Oka05} in diffuse clouds. The physical conditions of the     
\mbox{Sgr B2} envelope (\mbox{$T_{\rm k}$= 60 K} and \mbox{$n_{\rm H_2}$}= 10$^6$ cm$^{-3}$) are more similar 
to those of the molecular gas traced by our observations.\\
%
%_____________________________________________________________________________________________________________________________
%
\subsection{HN$^{13}$C chemistry.}
\label{Dis_hn13c}
\indent The chemistry of HNC is usually studied simultaneously with that of HCN. 
Unfortunately, we cannot compare \mbox{HN$^{13}$C} with \mbox{H$^{13}$CN} abundances in our sources. 
However, HCN and CS seem to share a constant abundance in regions dominated by different kinds of chemistries.
Both species are not strongly affected by shocks or by UV radiation. Therefore, we expect that the
\mbox{HN$^{13}$C/CS} and  \mbox{HN$^{13}$C/H$^{13}$CN} abundance ratios follow a similar trend. 
From \mbox{Table \ref{tab6}}, the lowest values of the \mbox{HN$^{13}$C/H$^{13}$CN} abundance ratio are associated 
with the sources that present the lowest values of the \mbox{HN$^{13}$C/CS} abundance ratio, the (20$''$, $-$20$''$) 
position of the Orion Bar and \mbox{IC443--G}. 
Conversely, the highest values of both ratios are found toward the (60$''$, $-$60$''$) position of the Orion PDR and 
the dark cloud TMC-1.
Following this trend, one can predict a lower \mbox{HN$^{13}$C/H$^{13}$CN} abundance ratio at the SW CND compared
with the 20 GMC.
This result agrees with previous studies that suggest that the HNC molecule is very sensitive to temperature, since it is
depleted toward regions of high temperatures caused by shock or UV-heating. \\
\indent In quiescent cool dark clouds, the \mbox{HNC/HCN} abundance ratio is close to or higher than unity 
(\mbox{2.1$\pm$1.2} in a range of \mbox{0.54--4.5}; \citealp{Hir98}). 
The \mbox{HNC/HCN} ratio decreases by \mbox{1--2} orders of magnitudes in the warmer GMCs 
near sites of massive star formation. Values ranging between \mbox{0.013--0.2} have been derived toward selected positions in 
the GMC OMC-1, with the lowest ratios found in the immediate vicinity of the hot core Orion-KL \citep{Sch92}.
These results suggest a correlation between the \mbox{HNC/HCN} abundance ratio and temperature.
Moreover, \citet{Hir98} claimed that this decrease in the ratio is due to destruction processes of HNC rather than 
to formation processes of HCN, as the HCN abundances in OMC-1 are almost the same than in dark cloud cores, whereas
the HNC abundances in OMC-1 are generally lower.
According to gas-phase chemical models, HCN and HNC are mainly produced by dissociative recombination of \mbox{HCNH$^+$},
leading to HNC on 60$\%$ and HCN on 40$\%$ of collisions, as observed abundances toward cold dark clouds
suggest \citep{Hir98}.
The dependence of HNC abundance on temperature has been explained by \citet{PDF90} as
due to neutral-neutral reactions with an activation barrier.
\citet{Hir98} derived an activation barrier of \mbox{190 K} for the destruction reactions of HNC with H and O.
However, this value could be higher according to quantum chemical calculations, making 
the destruction of HNC effective only in regions with temperatures higher than \mbox{300 K} \citep{TEH96}. \\
\indent As we can see in \mbox{Table \ref{tab3}}, the 20 GMC is the source that presents the lowest volume density of all our sources,
with \mbox{$n_{\rm H_2}$= 0.4$\times$10$^5$ cm$^{-3}$} at \mbox{$T_{\rm K}$= 50 K}. As previously mentioned,
a decrease in the temperature is compatible with an increase in the volume density. Thus, the 20 GMC could present 
a lower density or, alternatively, a lower temperature than the surrounding molecular gas, leading to less efficiency of 
the HNC destruction processes there. \\
%
%_____________________________________________________________________________________________________________________________
%
\subsection{The chemistry of the central 12 parsecs of the GC.}
\label{Dis_Sum}
\indent In summary, from the comparison of the abundance ratios derived toward the 12 central parsecs of the GC
with that of prototypical Galactic sources, one can clearly conclude that shocks are responsible
for maintaining the high SiO abundances found at the CND and the GMCs, while the strong UV radiation also affects 
the dense gas of the CND. This strong UV field seems to be responsible for the low abundances 
of HNCO and \mbox{HN$^{13}$C} measured toward the CND. Increasing temperatures due to UV-heating as one 
approaches the Central cluster could be responsible for the observed decrease in the \mbox{HN$^{13}$C} abundance.
The \mbox{HN$^{13}$C} peak abundance is found toward the well-shielded 20 GMC, a molecular cloud that could be either less dense 
or colder than the surrounding molecular gas. 
As shocks are widespread all over the central region, it is not expected that they have produced the differences found in the 
\mbox{HN$^{13}$C} abundances.
High temperatures can also favor HNCO formation in gas-phase. 
However, our observed abundance ratios toward the CND indicate that HNCO abundance is strongly depleted there, 
as well as in Galactic PDRs. The high SiO and HNCO abundance ratios found toward the 20 and 50 GMCs 
can be explained by the presence of nondissociative C-shocks with velocities between 20 and \mbox{30 km s$^{-1}$}, 
able to eject Si or SiO from grain mantles into the gas phase. 
HNCO could also be locked in grain mantles and directly realeased to the gas phase by these C-shocks.
Once in the gas phase, unshielded HNCO molecules would be rapidly destroyed by the UV radiation,
leading to the low HNCO abundances found at the CND molecular gas.  
The small gradient found in the \mbox{H$^{13}$CO$^+$} abundance from the CND to the 50 GMC could be explained 
by a decrease of one order of magnitude in the ionization rate with respect to the $``$standard$"$ value
toward this GMC, which presents the highest column density of all our sources.\\
%
%_____________________________________________________________________________________________________________________________
%
\subsection{The \mbox{HNCO(5--4)/SiO(2--1)} ratio as a probe of the distance to the nuclear region.}
\label{Dis_dist}
%
%_____________________________________________________________________________________________________________________________
%
\begin{figure*}
\centering
\includegraphics[width=16cm]{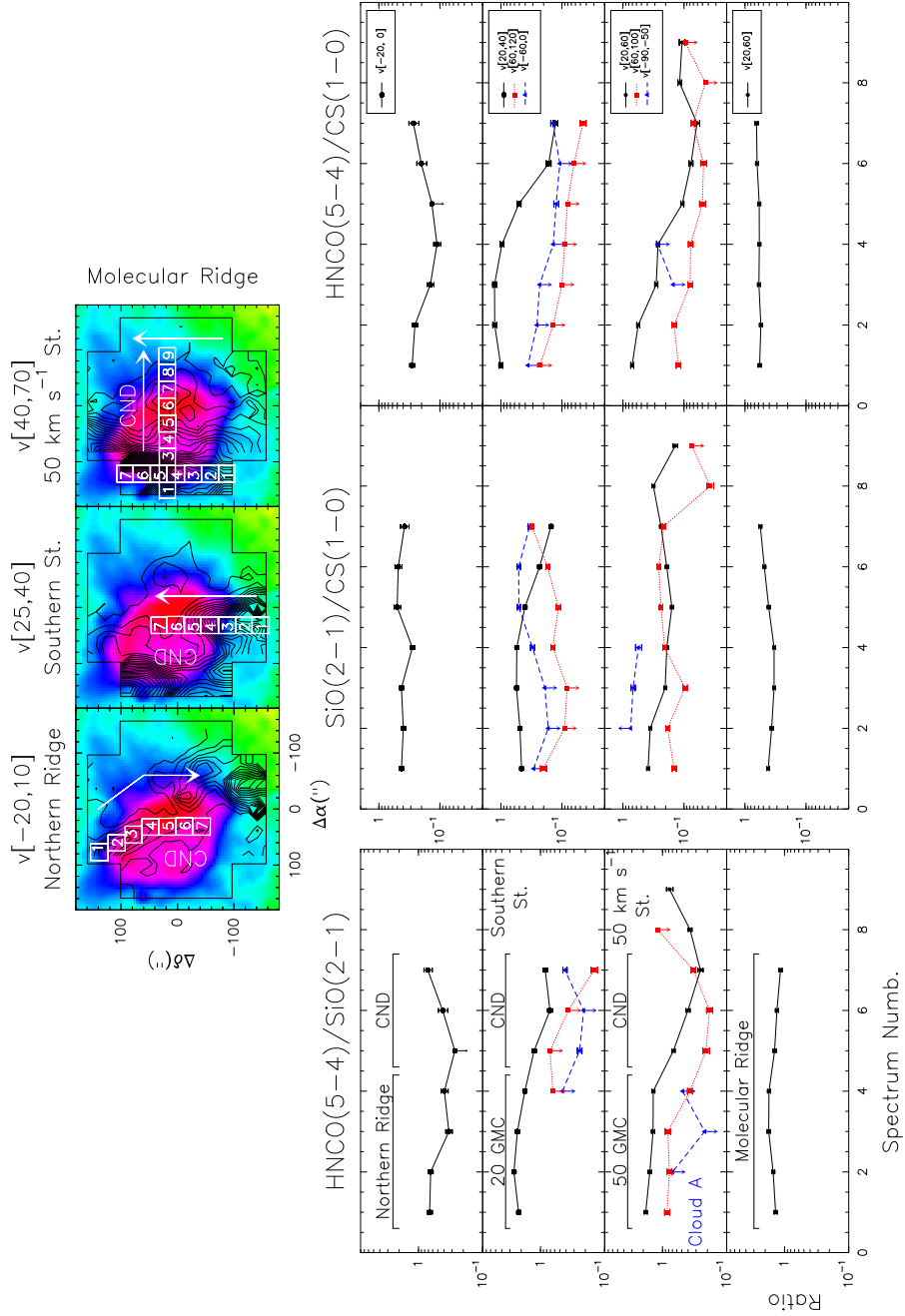}
\caption{\scriptsize{Velocity-integrated \mbox{HNCO(5--4)/SiO(2--1)} (left), 
\mbox{SiO(2--1)/CS(1--0)} (middle), and \mbox{HNCO(5--4)/CS(1--0)} (right) ratios along four 
different directions and different velocity intervals.
The four different directions are indicated at the top of the left panels, and 
they are pointed out with white arrows on the three maps located at the top of the figure.
These maps show, in contours, the \mbox{SiO(2--1)} emission in three different velocity ranges 
(indicated at the top of each map), superimposed on the radio continuum image 
of \citet{YZM87} at \mbox{20 cm} on a color scale.
The different velocity intervals in which we have integrated the molecular emission
are indicated at the upper-right corner of the right panels.
For example, three velocity-integrated ratios are shown along the Southern Streamer direction.
The one corresponding to the velocity interval of this streamer, \mbox{[20, 40] km s$^{-1}$}, is 
shown with black circles connected by black lines, whereas those corresponding to lower and higher velocity 
intervals, \mbox{[$-$60, 0]} and \mbox{[60, 120] km s$^{-1}$}, are represented by blue triangles 
connected by dashed lines and red squares connected by dotted lines, respectively.
The x-axis of the ratio plots represents the spectrum number, that increases from northeast to south in case of
the Northern Ridge (1$^{\rm st}$ plot beginning from the top), from south to north in the Southern Streamer (2$^{\rm nd}$), 
from east to west in the \mbox{50 km s$^{-1}$} Streamer (3$^{\rm rd}$), and from south to north in the Molecular Ridge
(4$^{\rm th}$), i.e., in the direction of the white arrows.
The white open squares, marked with the spectrum numbers, surround the region where the ratios have been derived, 
whose spectra can be seen in \mbox{Fig. \ref{fig11}}. 
Uncertainties in the ratio values represented by the errorbars have been calculated propagating 
the uncertainties in the intensities, which have been derived assuming a conservative 20$\%$ calibration 
error.}
\label{fig10}}   
\end{figure*}
%
%_____________________________________________________________________________________________________________________________
%
\begin{figure} 
\centering
\includegraphics[width=7cm]{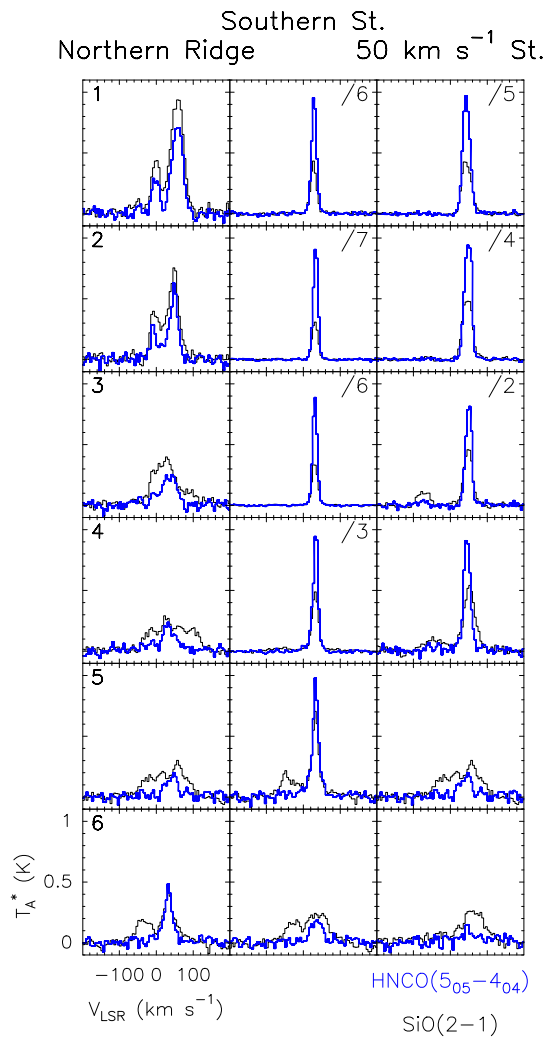}
\caption{\mbox{SiO(2--1)} (black line) and \mbox{HNCO(5--4)} (thick blue line) selected spectra taken along the 
Northern Ridge (left column), Southern Streamer (middle column) and \mbox{50 km s$^{-1}$} Streamer
(right column). 
In the maps at the top of \mbox{Fig. \ref{fig10}}, the areas where the spectra have been obtained
can be seen as white open squares. Their numbers appear in the upper-left corner of each 
spectrum of the left column.
The intensity of \mbox{SiO(2--1)} and \mbox{HNCO(5--4)} have been divided in some cases by a factor 
shown in the upper-right corner of the corresponding panel.
\label{fig11}}  
\end{figure}
%
%_____________________________________________________________________________________________________________________________
%
\indent The abundance ratios derived in \mbox{Sect. \ref{FracAbun}} show a possible relation between
the \mbox{HNCO/SiO} ratio and the distance to the nuclear region. 
The \mbox{HN$^{13}$C/H$^{13}$CO$^+$} and \mbox{HNCO/H$^{13}$CO$^+$} abundance ratios present a similar trend,
with the most contrasted column density ratios. 
However, \mbox{H$^{13}$CO$^+$} suffers from absorption
in the direction of \mbox{Sgr A$^*$}, whereas the \mbox{HN$^{13}$C} emission is not in general 
very intense. On the other hand, SiO and HNCO present strong emission in the regions where they are detected.
As we have previously discussed in \mbox{Sects. \ref{Dis_sio}} and \ref{Dis_hnco}, SiO and HNCO seem 
to share a common origin, that is, both molecules are locked in grain mantles and can be released to 
the gas phase by shocks. However, HNCO is photodissociated more efficiently than \mbox{SiO}.
In \mbox{Sect. \ref{FracAbun}} we have also seen that excitation effects do not modify the general trends
found in the integrated intensity ratio maps.
Therefore, the \mbox{HNCO/SiO} intensity ratio should provide information on the relative distance of the 
observed molecular gas to the Central cluster, the source of the UV radiation giving rise to the photodissociation 
in the nuclear region \citep{Kra95,Fig08}.\\    
\indent The \mbox{HNCO(5--4)} emission not only decreases in the high-velocity gas, but also in the velocity 
components related to the GMCs. \mbox{Figure \ref{fig10}} shows the \mbox{HNCO(5--4)/SiO(2--1)}, 
\mbox{SiO(2--1)/CS(1--0)}, and \mbox{HNCO(5--4)/CS(1--0)} intensity ratios 
(hereafter the \mbox{HNCO/SiO}, \mbox{SiO/CS}, and \mbox{HNCO/CS} ratios) for 
low, intermediate, and high-velocity gas ranges along four different directions. The Northern Ridge (velocity range 
between \mbox{[$-$20, 0] km s$^{-1}$}), the Southern Streamer ([$-$60, 0], [20, 40], and [60, 120] km s$^{-1}$), 
the \mbox{50 km s$^{-1}$} Streamer ([$-$90, $-$50], [20, 60], and [60, 100] km s$^{-1}$), and the  
Molecular Ridge ([20, 60] km s$^{-1}$). 
Maps at the top of \mbox{Fig. \ref{fig10}} indicate the spatially integrated regions 
where we derived the integrated instensity ratios.
In the following, we discuss each of this streamers comparing our instensity ratios with the 
previous pictures of the 3D arrangement proposed in the literature, trying to place them along the LOS:\\
\indent \textbf{$-$The Molecular Ridge:} 
The intensity ratio along these ridge is a good example of a rather constant \mbox{HNCO/SiO} 
ratio with only differences of a factor of 1.5 between the maximum and minimum observed values. 
This suggests that there the molecular gas is not strongly affected by UV radiation due to its distance 
to the UV source and/or its shielding from the UV photons.
The values of the \mbox{HNCO/SiO}, \mbox{SiO/CS}, and  \mbox{HNCO/CS} ratios are around 1.8, 0.6, and 0.6, 
respectively.\\
\indent \textbf{$-$The Northern Ridge:} 
This filamentary feature is located on the northern edge of \mbox{Sgr A East}, as can be seen from the 
\mbox{[$-$20, 10] km s$^{-1}$} map at the top-left of \mbox{Fig. \ref{fig10}}.
This ridge shows low \mbox{HNCO/SiO} ratios (\mbox{$<$0.7}) along all its full length 
(even for the more distant pixels from the nucleus), with a systematic decrease (by factor of \mbox{$\geq$2.5})
as we approach the center. 
This decrease must be due to the photodissociation of \mbox{HNCO}, as the \mbox{SiO/CS}
ratio shows a constant value along its length (\mbox{$\sim$0.4}). The \mbox{HNCO/CS} ratio is
low (0.11--0.3) compared to that of the Molecular Ridge and the GMCs.
It is likely that an even larger change in the \mbox{HNCO/SiO} ratio could be achieved along this filament
if the mapped area would have been extended further to the north of the ridge, as this feature seems to be connected with the 
\mbox{$-$30 km s$^{-1}$} molecular cloud \citep{SGu87}.
Our results support the idea that this cloud could be feeding the CND through the Northern Ridge.\\
\indent \citet{McG01} find a connection between this ridge and the northeastern region of the CND. 
These authors find a smooth positive velocity gradient in their \mbox{NH$_3$} maps from north 
to south (from \mbox{$-$10 km s$^{-1}$} in the Northern Ridge to \mbox{60 km s$^{-1}$} in the as a function
northeastern region of the CND). In the case that interaction between the ridge and the CND occurs, 
this velocity gradient places the ridge in front of the northeastern part of the CND. 
A similar gradient can be observed in the \mbox{declination-velocity} plot of the \mbox{SiO(2--1)} emission
averaged between \mbox{RA= [45$''$, 105$''$]} in \mbox{Fig. \ref{fig3}}. 
\citet{McG01} also find an increase in the \mbox{NH$_3$(2,2)/NH$_3$(1,1)} line ratios that suggests an 
increase in temperature where the streamer intersects the CND.\\ 
\indent Dust emission is also observed from the ridge, but it disappears in the connexion region between the CND 
and the ridge, which \citet{McG01} interpret as destruction of dust by UV-photons from the Central cluster. 
However, \citet{Lee08} suggest, from their kinematical comparison of shocked gas traced by \mbox{H$_2$} emission with 
quiescent gas traced by \mbox{NH$_3$} emission, that this feature is on the far side of the \mbox{Sgr A East} SNR, and is 
accelerated away from us due to the SNR expansion.
Both scenarios would be irreconcilable. In the case that the nuclear region were inside or in the leading edge of 
\mbox{Sgr A East} and the Northern Ridge were behind the SNR, the Northern Ridge could not be connected with the CND.
\citet{Kar03} have carried out an interferometric mapping of the \mbox{Sgr A} complex in the four \mbox{18--cm} OH 
absorption lines. The Northern Ridge is clearly seen in their 1665 and \mbox{1667 MHz} maps, providing a strong argument 
against the idea of \citet{Lee08}, suggesting that the ridge is in front of the \mbox{Sgr A East} SNR, i.e.: in front of
the continuum emission.
Our low \mbox{HNCO/SiO} ratio implies the Northern Ridge must be close to the nucleus, which supports the 
interaction scenario.\\
\indent \textbf{$-$The Southern Streamer:} 
This long filament extends northward from the \mbox{20 km s$^{-1}$} GMC toward the southeastern edge 
of the CND.
It has been observed in several transitions of \mbox{NH$_3$} 
\citep{Oku91,Ho91,Den93,CoH99,CoH00,McG01,HeH02,HeH05}, and in methanol \citep{Szc91}. 
The \mbox{HNCO/SiO} ratio along the Southern Streamer shows the largest gradient with a variation of 
a factor of \mbox{$\sim$18} between the ratio in the GMC (\mbox{$\sim$2.6} in the intermediate-velocity range) 
and in the CND (\mbox{$\sim$0.15} in the high-velocity range). 
Furthermore, the \mbox{HNCO/CS} ratio decreases by a factor of 28, with extreme values of
\mbox{$\sim$1.3} and \mbox{$\sim$0.05} in the GMC and the CND, respectively. 
This decrease in the ratio values as we move toward the center is also observed if we only consider  
the intermediate velocity range (a difference of a factor of 4 in the \mbox{HNCO/SiO} ratios), which could 
indicate that this streamer is approaching the central region and, therefore, a possible connection with the CND.
Similar to the Northern Ridge, this decrease in the ratio is probably caused by the photodissociation of the \mbox{HNCO}, 
together with the enhancement of the \mbox{SiO} abundance in the CND, combined with a less effective photodissociation 
rate of \mbox{HNCO} in the GMC. 
The \mbox{SiO/CS} ratio decreases by a factor of 4 as we approach the center
in the intermediate velocity range, whereas the \mbox{HNCO/CS} ratio decreases in a 
factor of 10.
The \mbox{SiO} and \mbox{HNCO} line profiles for the \mbox{20 km s$^{-1}$} GMC are very similar, peaking at the same 
velocities as one approaches the center.
However, the high-velocity gas of the CND only appears in the SiO emission 
(see the middle column of \mbox{Fig. \ref{fig11}}).\\
\indent No velocity gradient has been detected along the Southern Streamer \citep{Ho91}; therefore, if gas is accelerated 
along it, it must occur mainly perpendicular to the LOS, which complicates the connection between the CND and
the \mbox{20 km s$^{-1}$} GMC, since the GMC is located in front of the nucleus \citep{GuD80,Par04}. 
\citet{MCa09} have compared the \mbox{HCN(4--3)/HCN(1--0)} intensity ratio map of the central region, 
which traces highly excited material in the CND, with the \mbox{NH$_3(3,3)$} and \mbox{(6,6)} observations 
\citep{HeH05}, which trace the Southern Streamer and very warm gas, respectively. 
They find that at the location where the northeast part of the Southern Streamer 
and the southeastern part of the CND meet, the strongest peak in the \mbox{NH$_3(6,6)$} emission and
a region of high \mbox{HCN(4--3)/HCN(1--0)} ratio are found. Therefore, they propose that the impact of
the Southern Streamer with the CND caused the infall of material toward the black hole, which is responsible
for the highly excited gas found as we approach the center.  
Additionally, \citet{Sat08}, from interferometric observations, find an increase in the 
\mbox{SiO(2--1)/H$^{13}$CO$^+$(1--0)} ratio at the possible interaction site between the southeastern 
region of the CND and the Southern Streamer, together with larger line widths.
Another probe for shocks in this possible interaction site is the \mbox{36 GHz} methanol maser 
recently found by \citet{SPF10} (see their \mbox{Fig. 2}) with a velocity of \mbox{22 km s$^{-1}$} (these Class I 
masers are collisionally pumped and considered as typical signposts of shock-excited material).
Moreover, \citet{Lee08} claim that the SNR \mbox{Sgr A East} drives shocks into the northernmost part of the 
Southern Streamer and the southern part of the CND, pushing the gas toward us.
These interactions with the SNR make the connection between the Southern Streamer and the CND possible.
Therefore, the decrease in the \mbox{HNCO/SiO} ratio as the Southern Streamer approaches 
the nuclear region confirms the interaction between the Southern Streamer and the CND.\\   
\indent \textbf{$-$The \mbox{50 km s$^{-1}$} Streamer:} 
This ridge of gas extending westward from the \mbox{50 km s$^{-1}$} GMC was first
detected by \citet{Szc91} in methanol and was also observed in \mbox{HCN(3--2)} by \citet{Mas95}. 
This streamer could be a possible connection between the \mbox{50 km s$^{-1}$} GMC and the northern 
lobe of the CND. 
Our data show a decrease in the \mbox{HNCO/SiO} ratio that would indicate that this streamer 
is approaching the nucleus; however, if we compare the SiO(2--1) and HNCO(5--4) spectra extracted from
the possible interaction region between the streamer and the CND (spectrum number 4 in the right column 
in \mbox{Fig. \ref{fig11}}), it seems that both molecules are tracing different gas components. 
The \mbox{HNCO(5--4)} line is tracing the \mbox{50 km s$^{-1}$} GMC with very 
similar profiles to spectra with numbers $<$4, which sample the inner region of this molecular cloud. 
All of these \mbox{HNCO} profiles do not show line widths that are too broad, \mbox{$\sim$(19--23) km s$^{-1}$}. 
In contrast, the \mbox{SiO(2--1)} lines taken at the same positions show an increase in the line widths from 
\mbox{$\sim$(27--31) km s$^{-1}$} (spectra with numbers \mbox{$<$4}) to \mbox{37 km s$^{-1}$} (spectrum number 4). 
Moreover, the central velocity of the \mbox{SiO(2--1)} line is slightly redshifted with respect to the \mbox{HNCO(5--4)} line 
in spectrum number 4.\\
\indent \citet{Mas95} do not detect any velocity gradient along the feature, and considering 
that the GMC is located behind the SNR \mbox{Sgr A East}, the connection is 
unlikely unless part of this GMC lies in the same plane as the northern lobe of the CND. In that case, the gas 
would run perpendicular to the LOS and no velocity gradient would be expected \citep{Mas95}.
They suggest that this apparent connection could also be due to the overlapping of material from the GMC and 
the CND, as their \mbox{HCN(4--3)} maps suggest.
Therefore, in this case, the decrease in the \mbox{HNCO/SiO} ratio could be produced by the blending of 
two gas components and not to a real interaction: the \mbox{50 km s$^{-1}$} GMC, where the \mbox{HNCO} is more prominent 
than the \mbox{SiO} line, and gas from the CND, where the \mbox{HNCO} line disappears because of photodissociation from the central 
UV radiation and \mbox{SiO} emission is broadened by shocks.\\  

There are two other features, Cloud A and the Western Streamer, that do not seem to be connected with the CND, 
but their low \mbox{HNCO/SiO} ratio indicates that they are close to the nuclear region. 
Both features seem to be related to the SNR \mbox{Sgr A East}:\\ 
\indent \textbf{$-$Cloud A:} 
This cloud was previously observed by \citet{Gen90}
in \mbox{C$^{18}$O(2--1)}, and mapped by \citet{Sut90} in \mbox{CO(3-2)} 
and \citet{Ser92} in \mbox{CS(5--4)} and \mbox{CS(7--6)}. 
In addition, \citet{Ser92} found a redshifted emission (\mbox{80--85 km s$^{-1}$}) just  
to the south of Cloud A. Both high-velocity emission components
lie just inside of the compressed ridge of the \mbox{50 km s$^{-1}$} GMC, which suggests that  
the molecular gas of this GMC had been accelerated to very high
velocities by the expansion of \mbox{Sgr A East}. The accelerated molecular gas appears on the
front and the back sides of the expanding radio shell. \citet{Wri01} support this scenario from  
their interferometric \mbox{HCO$^+$} and \mbox{HCN} maps, together with
other near blue and redshifted clumps of gas inside of the compressed ridge.
A different interpretation was proposed by \citet{Lis95}, who detected this feature with very strong 
emission in \mbox{CO(3--2)} at \mbox{v$\leq-$40 km s$^{-1}$}, 
but they considered this feature as an extension of the high-longitude forbidden-velocity part of the CND, 
as they see a connection between it and the emission arising from the southwestern region of the CND,
concluding that Cloud A was not an isolated feature.\\ 
\indent From the declination-velocity cut made through it (panel \mbox{RA=[45$''$, 105$''$]} in 
\mbox{Fig. \ref{fig8}}), this cloud does not appear to be kinematically 
connected to either the \mbox{50 km s$^{-1}$} GMC or to any other gas component.
The \mbox{SiO/CS} and \mbox{HNCO/SiO} ratios in this cloud ($\sim$ 0.7 and \mbox{$\leq$ 0.22}, 
respectively; see \mbox{Fig. \ref{fig10}}) suggest there are shocks, as well as UV-radiation,
which also explain the lack of \mbox{NH$_3$} and \mbox{1.2 mm} dust emission, since the ammonia molecules and
dust grains are destroyed by photodissociation and shocks, respectively (see \mbox{Fig. 9} of \citealt{McG01}).
Shocks could be caused by the expansion of \mbox{Sgr A East} SNR into this cloud, whereas UV-radiation could 
originat in the Central cluster, locating this cloud close to the nuclear region.
The \mbox{SiO/CS} ratio in Cloud A is one of the highest found in the region, only exceeded by
the southwest lobe of the CND. This fact could be explained by the geometry of the shock, that could be almost along the LOS.
Additional support for the SNR interaction scenario comes from the central velocities of the \mbox{SiO(2--1)} and \mbox{CS(1--0)} 
profiles.
The \mbox{SiO(2--1)} central velocity is offset by \mbox{$-$11$\pm$3 km s$^{-1}$} with respect 
to that of \mbox{CS(1--0)}, indicating that Cloud A is pushed toward us by the SNR \mbox{Sgr A East}, if we
consider that SiO traces more efficiently the shocked gas, whereas \mbox{CS} traces more quiescent gas.
Moreover, Cloud A is detected in the \mbox{18--cm} OH absorption maps of \citet{Kar03} (in the 1665 and \mbox{1667 MHz} lines),
which locates it in front of the continuum emission of the SNR, consistent with our finding\\ 
\indent \textbf{-The Western Streamer:} This feature is a long curved filamentary structure that closely matches  
the western edge of \mbox{Sgr A East}. \citet{McG01} reports a velocity gradient of \mbox{1 km s$^{-1}$ arcsec$^{-1}$} 
along its \mbox{150$''$} length. The southern half has a negative velocity whereas the northern half has a positive velocity,
consistent with a ridge of gas highly inclined to the LOS that is pushed outward by the expansion of the SNR.
The northern part of the streamer would be located on the far side of the SNR producing the redshifted emission,
whereas the southern part would be located on the front side of the shell producing the blueshifted velocities 
\citep{McG01,Lee08}.
The \mbox{SiO/CS} ratios in this streamer present intermediate values of \mbox{0.3$-$0.4}. 
On the other hand, the \mbox{HNCO/SiO} ratios seem to be low (0.3$-$0.4) with very low-intensity 
emission of \mbox{HNCO}, just slightly over the rms noise.
This enhancement in the \mbox{SiO/CS} ratios could be attributed to shocks caused by the 
expansion of the SNR \mbox{Sgr A East} into the molecular filament. The smaller enhancement of the ratio with respect to
the enhancement observed in the southwestern lobe of the CND and Cloud A could be due to the geometry of the interaction.
In this case it must be highly inclined and orthogonal to the LOS. 
The Western Streamer is not detected in the 1.2 mm dust emission map (see \mbox{Fig. 9} of \citealt{McG01}). 
These authors propose that the gas of the streamer could originate closer to the nucleus, where most of the dust is 
destroyed by UV-photons from the central cluster, and then pushed outwards by the SNR. 
Our data suggest that the dust destruction is very likely caused by shocks.
Molecular gas of the Western Streamer also presents the highest kinetic temperatures in the central \mbox{10 pc}, even showing 
\mbox{NH$_3$(6,6)} emission that traces gas at \mbox{$\geq$100 K} \citep{HeH05}.\\ 

\indent In summary, the \mbox{HNCO/SiO} intensity ratio seems to be an excellent tool for locating the different gas 
components relative to the nucleus. While the \mbox{50 km s$^{-1}$} Streamer seems to be due to the overlapping 
of two different gas components (CND $+$ \mbox{50 km s$^{-1}$} GMC gas), the Southern Streamer seems to be feeding the 
CND from the \mbox{20 km s$^{-1}$} GMC, whereas the Northern Ridge remains a good candidate to also feed the CND, 
possible connecting the \mbox{$-$30 km s$^{-1}$} molecular cloud with the central region.
The Western Streamer and Cloud A seem to be associated with the expansion of the SNR \mbox{Sgr A East} and located close 
to the nuclear region. 
This streamer would arise from the western border of the SNR interacting with the molecular gas, whereas Cloud A could be 
part of the \mbox{50 km s$^{-1}$} GMC that has been pushed by the SNR toward us almost along the LOS.\\
%
%_____________________________________________________________________________________________________________________________
%
\section{Conclusions}
\label{Concl}
We have presented maps of the central \mbox{12 pc} of the GC in the \mbox{$J =$ 2 $\rightarrow$ 1} 
transition of \mbox{SiO}, the \mbox{$J =$ 5$_{0,5}$ $\rightarrow$ 4$_{0,4}$} transition of \mbox{HNCO}, and
the \mbox{$J =$ 1 $\rightarrow$ 0} transition of \mbox{H$^{13}$CO$^+$},
\mbox{HN$^{13}$C}, and \mbox{C$^{18}$O}, observed with the IRAM-30m telescope at Pico Veleta,
with an angular resolution of \mbox{30$''$}. \\
\indent The main results obtained from this work can be summarized as follow: 
\begin{itemize}
 \item The \mbox{SiO(2--1)} emission, observed in a velocity range of \mbox{[$-$125, 130] km s$^{-1}$}, 
       perfectly traces the CND, as well as all the features previously identified in other high-density 
       molecular tracers. 
       The \mbox{H$^{13}$CO$^+$(1--0)} emission covers the same velocity range and follows the SiO 
       morphology, except for the absorption features toward the radio continuum source \mbox{Sgr A$^*$}.
 \item In contrast, the \mbox{HNCO(5--4)} molecular line shows emission in the narrowest velocity range
       (\mbox{[$-$30, 85] km s$^{-1}$}). 
       This molecular line presents the strongest emission of all the observed lines; however, its emission is 
       concentrated in the GMCs and the Molecular Ridge, quickly disappearing as one approaches the Central cluster.
       Similar behavior is shown by the \mbox{HN$^{13}$C(1--0)} emission. 
 \item The \mbox{C$^{18}$O(1--0)} emission is affected by contamination with foreground gas due to its lower critical
       density.
 \item Emission mostly coming from \mbox{NH$_2$CHO(5$_{1,4}-$4$_{1,3}$)} has been detected toward the GMCs and 
       the Molecular Ridge. 
 \item Using the LVG approximation, volume densities of \mbox{$\sim$10$^5$ cm$^{-3}$} assuming \mbox{$T_{\rm K}$= 50 K} 
       are derived for all the gas components, expect for the the \mbox{20 km s$^{-1}$} GMC, which could be 
       either less dense or colder.
       In general, the highest densities are found toward the CND and Western Streamer, and the lowest ones 
       toward the two GMCs.
 \item Hydrogen column densities have been derived from the high-density tracer CS, obtaining values of
       \mbox{$0.5-6\times10^{22} {\rm cm^{-2}}$}, with the highest value found toward the 
       \mbox{50 km s$^{-1}$} GMC.      
 \item The HNCO molecule shows the strongest contrast in column density and fractional abundance, 
       with factors of \mbox{$\geq$60} (\mbox{$\leq0.8-50\times10^{13} {\rm cm^{-2}}$}) and 
       28 (\mbox{$\sim0.7-19\times10^{-9}$}), respectively.
       The highest values are found toward the cores of the GMCs, whereas the lowest ones are derived toward 
       the positions associated with the CND. 
       A similar behavior is also found for \mbox{HN$^{13}$C}, with abundances \mbox{$\sim0.06-1.3\times10^{-9}$}.
\item  SiO abundances do not follow the HNCO trend (\mbox{$0.20-2.4\times10^{-9}$}), with high values at the cores 
       of the GMCs, as well as toward the CND, whereas \mbox{H$^{13}$CO$^+$} shows the smallest variation in its fractional 
       abundance (\mbox{$0.05-0.4\times10^{-9}$}). 
       Thus, the observed species can be grouped into those whose abundances decrease (HNCO and \mbox{HN$^{13}$C}), 
       and those whose abundances remain similar or are enhanced toward the nuclear region (SiO and \mbox{H$^{13}$CO$^+$}).
 \item From the comparison of the abundances derived toward the 12 central parsecs of our Galaxy with 
       those of prototypical Galactic sources, one can conclude that non-dissociative C-shocks
       with velocities between \mbox{20--30 km s$^{-1}$}, able to eject Si/SiO from the grain mantles to the gas phase,
       are responsible for maintaining the high SiO abundances found in the CND and the GMCs.
       HNCO could also be locked in grain mantles and directly realeased to the gas phase by the C-shocks.  
       However, once in the gas phase, HNCO molecules are not well-shielded from the UV radiation coming from 
       the Central cluster and would be rapidly destroyed, leading to the low abundances found in the CND.
       This strong UV field also seems to be responsible for the low abundances of \mbox{HN$^{13}$C} measured 
       toward the CND, as high temperatures could favor neutral-neutral reactions leading to the destruction of \mbox{HNC}
       molecules.
       The small gradient found in the \mbox{H$^{13}$CO$^+$} abundance from the CND to the \mbox{50 km s$^{-1}$} GMC could 
       be explained by a decrease of one order of magnitude in the ionization rate with respect to the $``$standard$"$ value
       toward the \mbox{50 km s$^{-1}$} GMC.
 \item Ratios between \mbox{SiO} (which traces shocked gas), \mbox{HNCO} 
       (shocked and/or photodissociated gas), and \mbox{CS} (quiescent dense gas) emissions provide
       an excellent tool for studing the possible connections and the 3D arrangement of the different 
       features found in the central region of the Galaxy. 
       From their comparison, we conclude that not only is the CND affected by 
       UV radiation, but also the Western Streamer and Cloud A. 
       The Southern Streamer and the Northern Ridge could be feeding the 
       CND; however, the \mbox{50 km s$^{-1}$} Streamer, composed of two overlapping components, 
       does not seem to be connected with the CND.\\          
\end{itemize}
%
%_____________________________________________________________________________________________________________________________
%
\begin{acknowledgements}
The authors would like to thank Prof. M. Tsuboi and D. M. Montero-Casta\~{n}o for kindly providing their
CS(1--0) and HCN(4--3) data, respectively. 
We also thank the referee for the suggestions. 
This work was supported by the Spanish Ministerio de Ciencia e Innovaci\'on under 
project \mbox{ESP2007-65812-C02-01}. 
\end{acknowledgements} 
%
%_____________________________________________________________________________________________________________________________
%
\bibliographystyle{aa} 
\bibliography{aa.bib}
%
%_____________________________________________________________________________________________________________________________
%
\Online
%
%_____________________________________________________________________________________________________________________________
%
\begin{appendix} %First online appendix
\section{Velocity channel maps}
\label{ApendixA}
%
%_____________________________________________________________________________________________________________________________
%
\begin{figure*}
\centering
\includegraphics[width=16cm,clip]{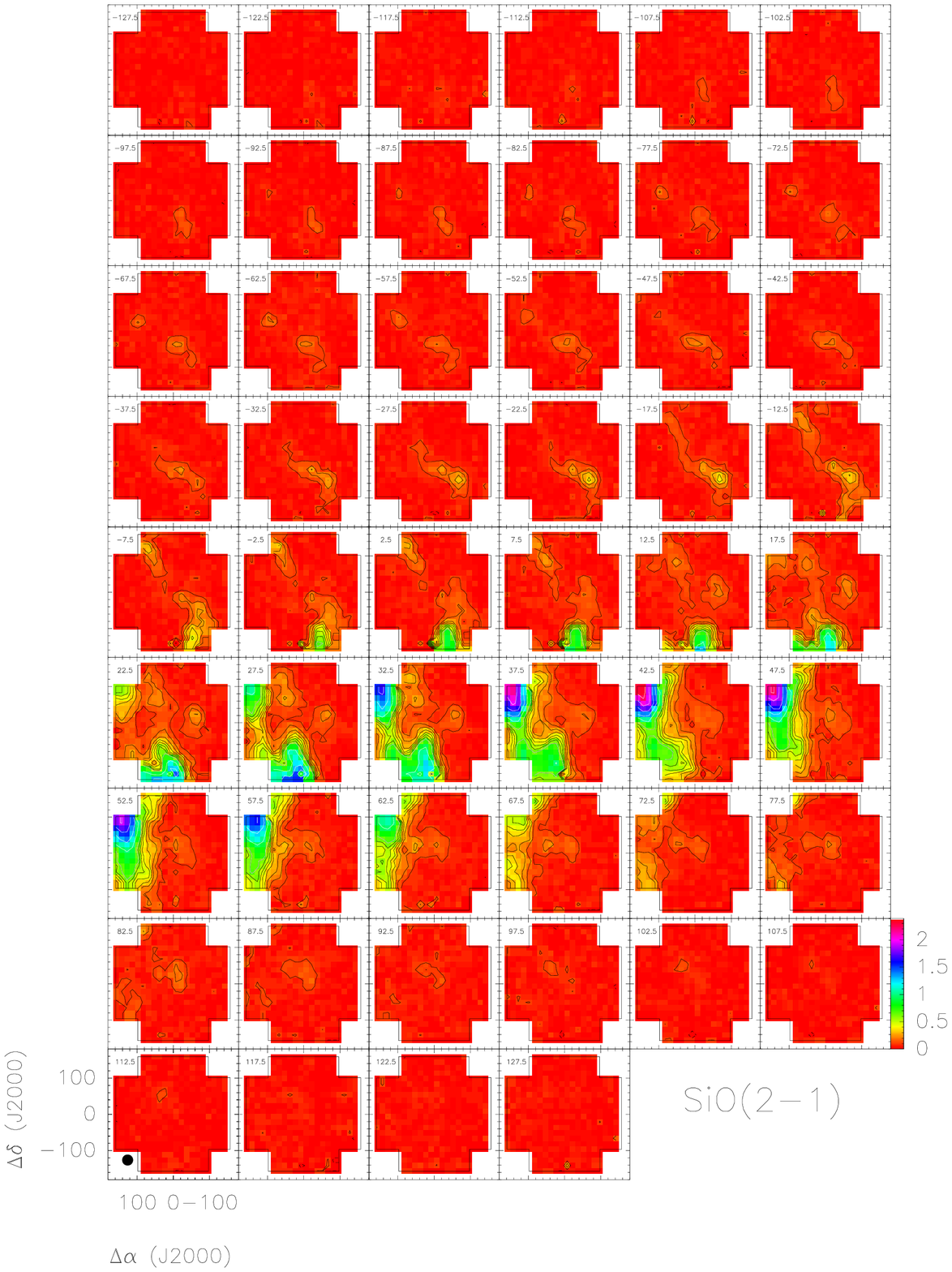}
\caption{\scriptsize{Velocity-channel maps of the \mbox{SiO(2--1)} emission.
Black contour levels for the molecular line emission (in the \mbox{$T_{\rm A}^*$} scale) 
are from \mbox{0.08 K} (\mbox{3$\sigma$}) to \mbox{0.64 K} in steps of \mbox{0.08 K}. 
White contour levels are from \mbox{0.64 K} in steps of \mbox{0.32 K}.
Dashed countours correspond to the \mbox{$-$3$\sigma$} level.
Velocity channels range from \mbox{$-$130} to \mbox{130 km s$^{-1}$} by steps of 5 km s$^{-1}$ 
(the velocity width per channel).
The central velocities of the velocity-channel maps are shown in the upper-left corner of each panel. 
The wedge at the side shows the intensity scale of the \mbox{SiO(2--1)} emission.
The beam size (30$''$) is shown at the bottom-left corner.
\mbox{Sgr A$^*$} is the origin for the offset coordinates (in arcseconds).} 
\label{appfig1}}
\end{figure*}
%
%_____________________________________________________________________________________________________________________________
%
\begin{figure*}
\centering
\includegraphics[width=16cm,clip]{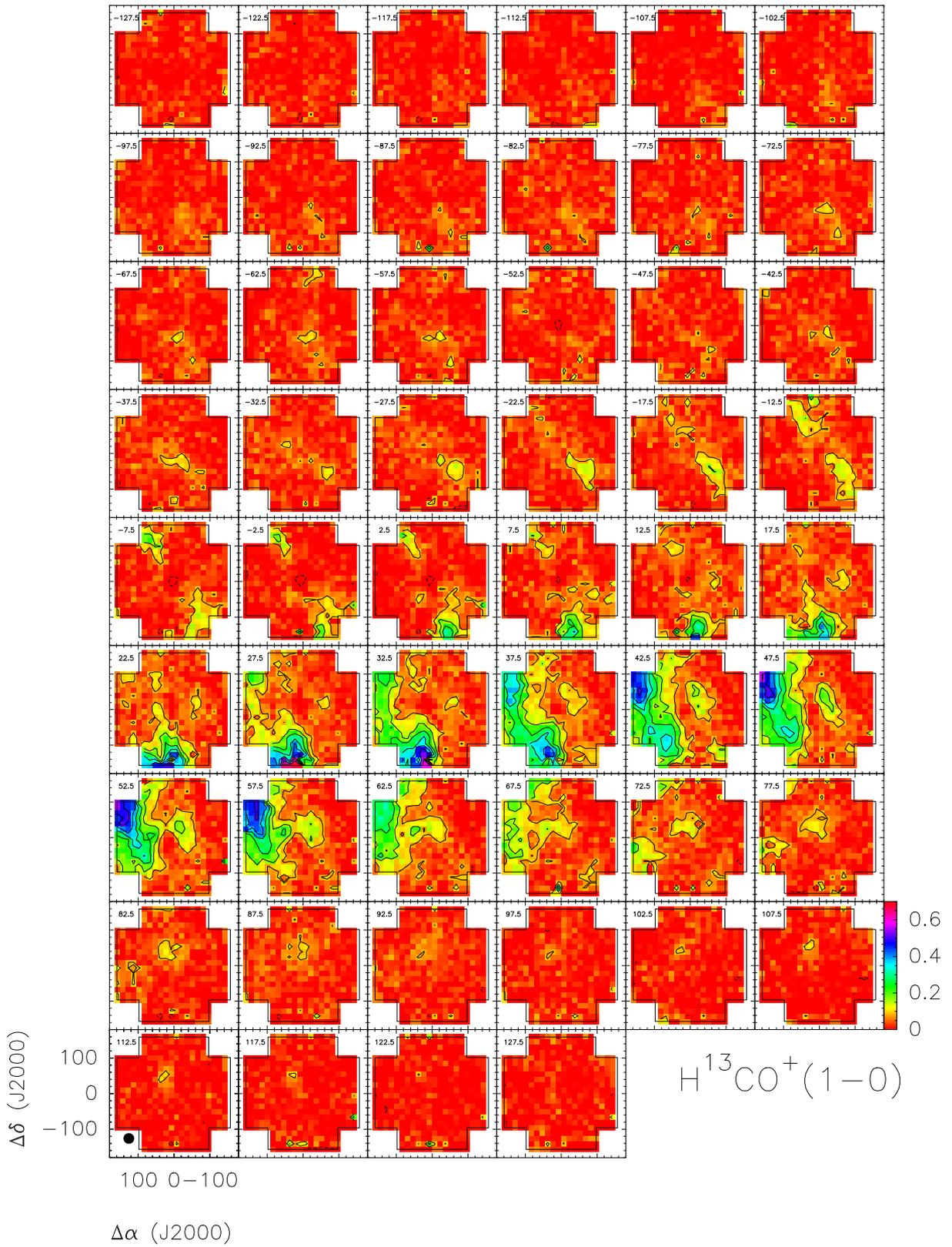}
\caption{\scriptsize{Velocity-channel maps of the \mbox{H$^{13}$CO$^+$(1--0)} emission.
Contour levels for the molecular line emission are from \mbox{0.08 K} (\mbox{3$\sigma$}) 
in steps of \mbox{0.08 K}. 
The other characteristics of the figure are the same as in \mbox{Fig. \ref{appfig1}}.}
\label{appfig2}}  
\end{figure*}
%
%_____________________________________________________________________________________________________________________________
%
\begin{figure*}
\centering
\includegraphics[width=16cm,clip]{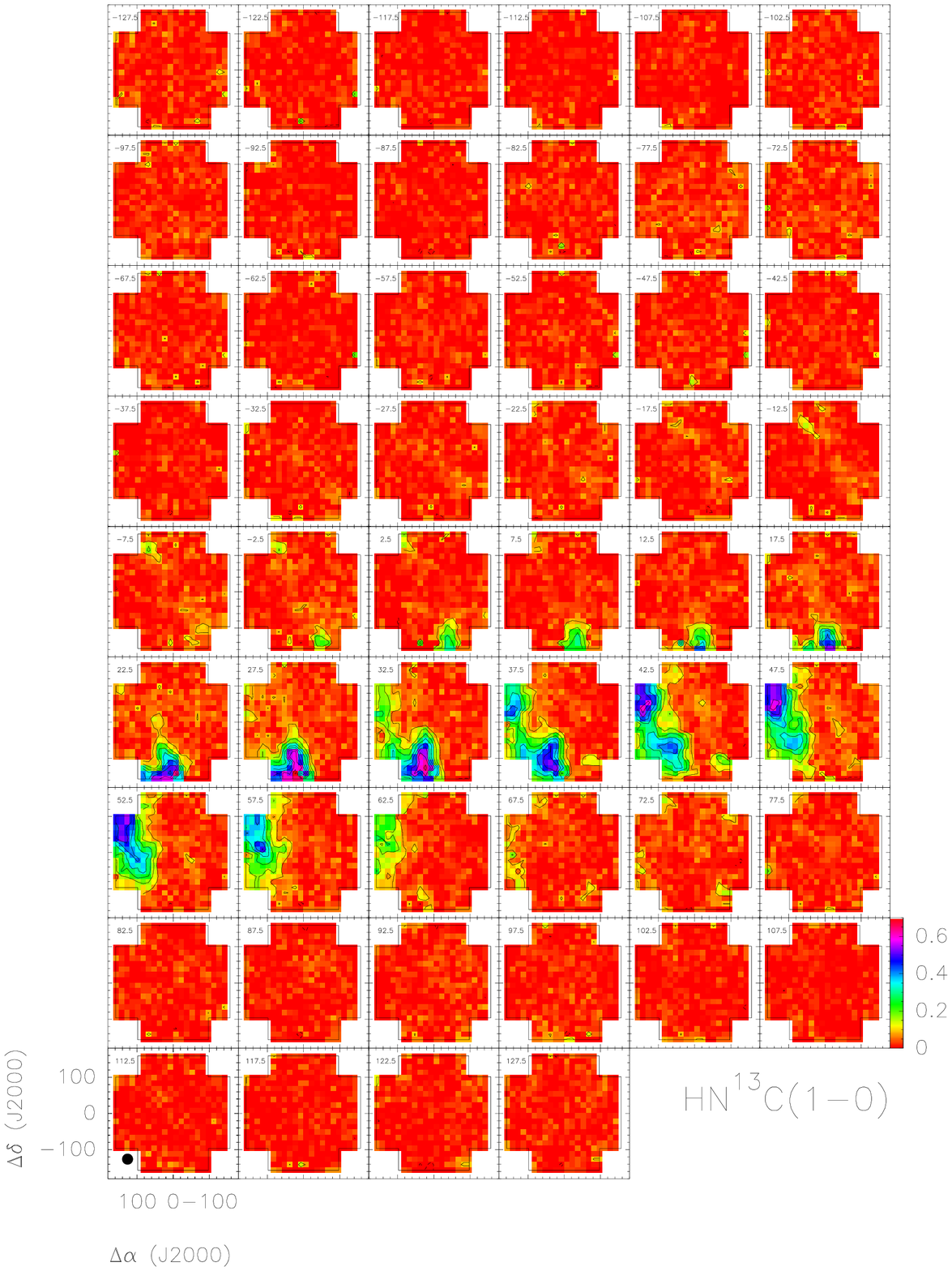}
\caption{\scriptsize{Same as \mbox{Fig. \ref{appfig2}} but for the \mbox{HN$^{13}$C(1--0)} emission.}
\label{appfig3}}  
\end{figure*}
%
%_____________________________________________________________________________________________________________________________
%
\begin{figure*}
\centering
\includegraphics[width=16cm,clip]{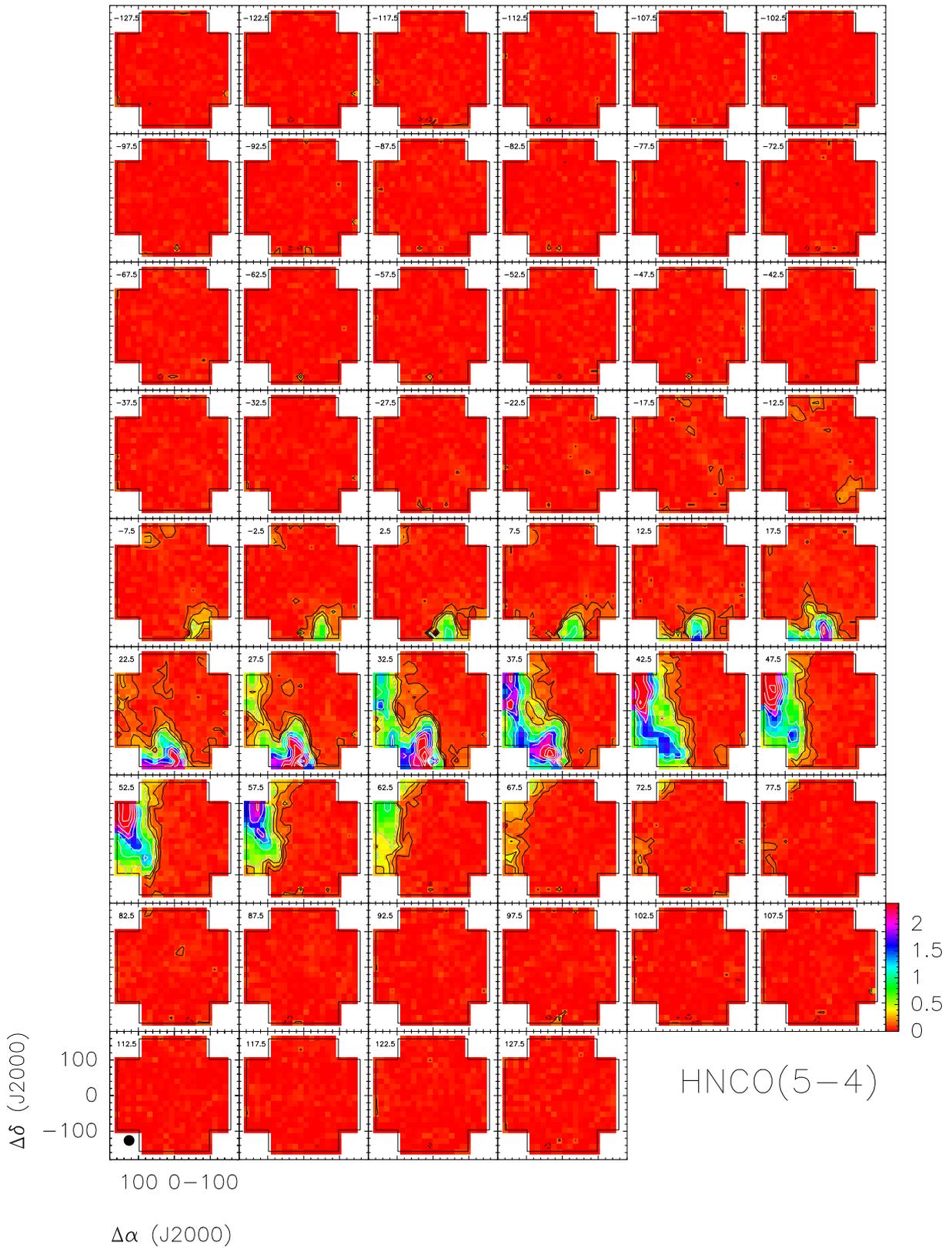}
\caption{\scriptsize{Velocity-channel maps of the \mbox{HNCO(5$_{0,5}$--4$_{0,4}$)} emission.
Black contour levels for the molecular line emission are from \mbox{0.11 K} (\mbox{3$\sigma$}) 
to \mbox{0.44 K} in steps of \mbox{0.11 K}. 
White contour levels are from \mbox{0.44 K} in steps of \mbox{0.44 K}.
In order to show more clearly the spatial distribution of the weak emission, 
the intensity scale has been saturated.
The other characteristics of the figure are the same as in \mbox{Fig. \ref{appfig1}}.}
\label{appfig4}}  
\end{figure*}
%
%_____________________________________________________________________________________________________________________________
%
\begin{figure*}
\centering
\includegraphics[width=16cm,clip]{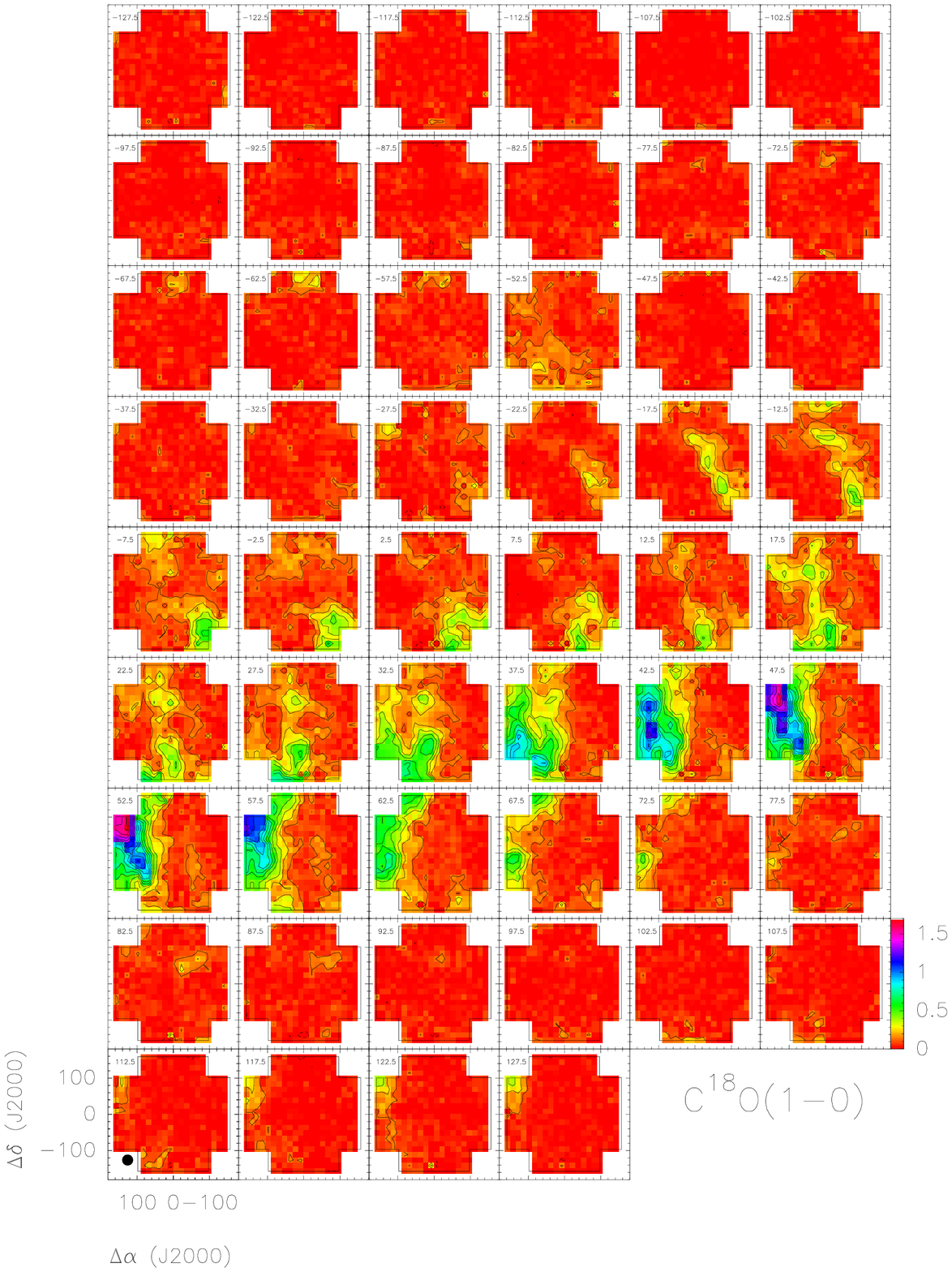}
\caption{\scriptsize{Velocity-channel maps of the \mbox{C$^{18}$O(1--0)} emission.
Contour levels for the molecular line emission are from \mbox{0.11 K} 
(\mbox{3$\sigma$}) in steps of \mbox{0.11 K}. 
The other characteristics of the figure are the same as in \mbox{Fig. \ref{appfig1}}.}
\label{appfig5}}  
\end{figure*}
%
%_____________________________________________________________________________________________________________________________
%
\end{appendix}
%
%_____________________________________________________________________________________________________________________________
%
\begin{appendix} %Second online appendix
\label{ApendixB}
\section{Integrated intensity ratio maps}
%
%_____________________________________________________________________________________________________________________________
%
\begin{figure*}
\centering
\includegraphics[width=15cm,clip]{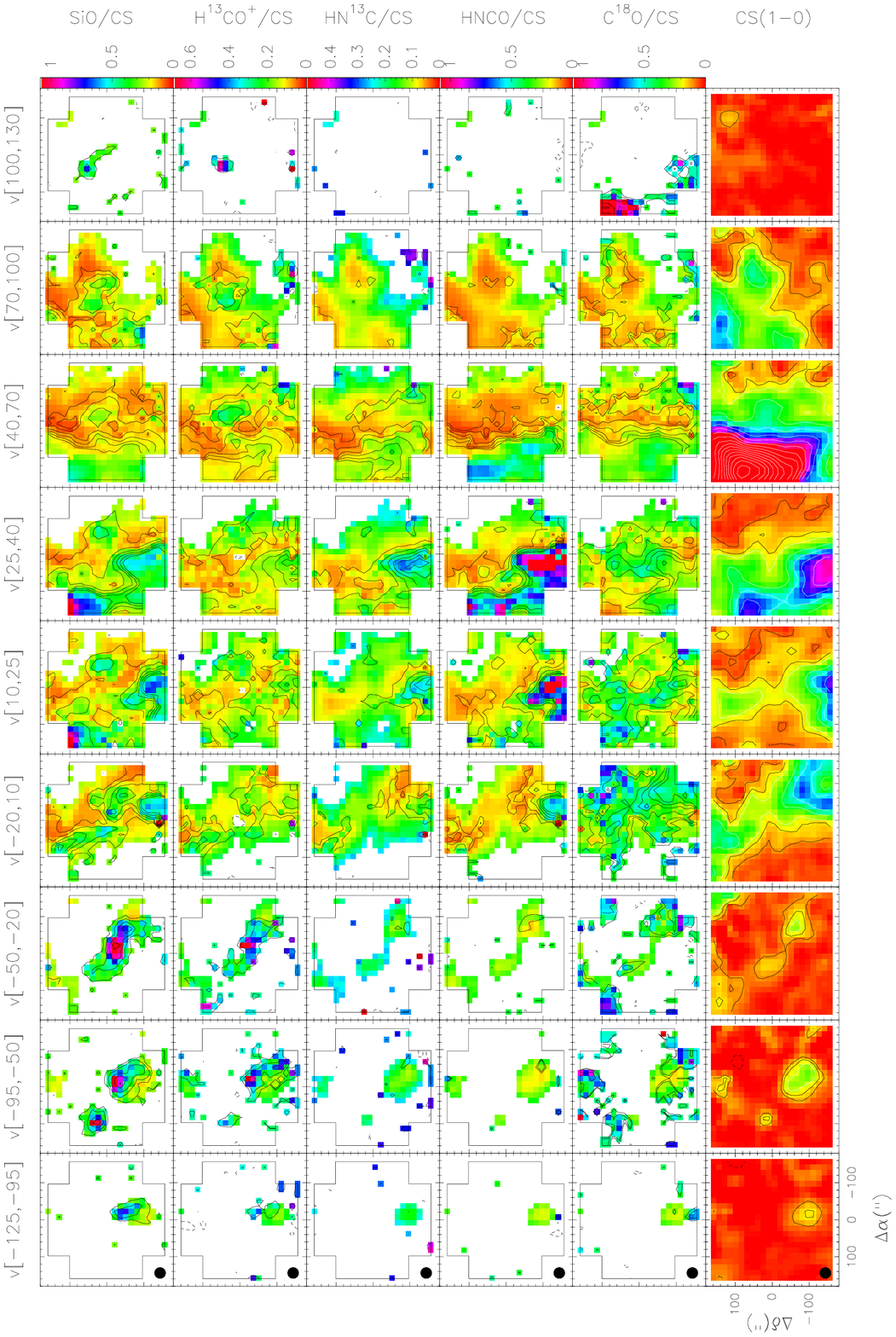}
\caption{\scriptsize{\mbox{X/CS(1--0)} intensity ratios in the same selected velocity ranges as \mbox{Fig. \ref{fig2}}.
X stands for \mbox{SiO(2--1)}, \mbox{H$^{13}$CO$^+$(1--0)}, \mbox{HN$^{13}$C(1--0)}, \mbox{HNCO(5--4)}, and \mbox{C$^{18}$O(1--0)}.
Molecular emission is shown in contour levels, which are \mbox{$-$3$\sigma$} (dashed contour), \mbox{3$\sigma$}, 
from 2.0 to 8 in steps of \mbox{1.5 K km s$^{-1}$}.
The \mbox{3$\sigma$} level of each molecule is the same as in \mbox{Fig. \ref{fig2}}. 
Intensity ratio maps take upper and lower limits into account, using the \mbox{3$\sigma$} value of each map as
the limit value.
The wedges to the left show the color scale of the different intensity ratios.
Row at the bottom shows \mbox{CS(1--0)} maps of \citet{Tsu99} to compare and show which pixels present 
meaningful ratios or are only limits (their contour levels and intensity scale are the same as in \mbox{Fig. \ref{fig2}}).}
\label{appfig6}}  
\end{figure*}
%
%_____________________________________________________________________________________________________________________________
%
\end{appendix}
%
%_____________________________________________________________________________________________________________________________
%
\end{document}